\documentclass[twocolumn]{revtex4}
\usepackage{graphicx}
\usepackage[]{epsfig}
\usepackage{bm}
\usepackage{color}

\newcommand{\beq}{\begin{equation}}
\newcommand{\eeq}{\end{equation}}
\newcommand{\beqd}{\begin{displaymath}}
\newcommand{\eeqd}{\end{displaymath}}
\newcommand{\beqa}{\begin{eqnarray}}
\newcommand{\eeqa}{\end{eqnarray}}

\renewcommand{\a}{\alpha}

\newcommand{\sign}{{\rm sign}}

\newcommand{\arctanh}{{\rm arctanh}}
\newcommand{\comment}[1]{}

\newcommand{\Tr}{{\rm Tr}\,}

\newcommand{\hR}{\hat{R}}
\newcommand{\hT}{\hat{T}}
\newcommand{\hX}{\hat{X}}
\newcommand{\hx}{\hat{x}}
\newcommand{\hmu}{\hat{\mu}}
\newcommand{\Tau}{\mathcal{T}}

\begin{document}

\title{Path Integral Approach  Unveils the Role of Complex Energy Landscape for Activated Dynamics of Glassy Systems}
\author{Tommaso Rizzo}
\affiliation{ISC-CNR, UOS Rome, Universit\`a ``Sapienza'', Piazzale A.~Moro 2, I-00185, Rome, Italy}
\affiliation{Dip.\ Fisica, Universit\`a ``Sapienza'', Piazzale A.~Moro 2, I--00185, Rome, Italy}

\begin{abstract}
The complex dynamics of an increasing number of systems is attributed to the
emergence of a rugged energy landscape with an exponential number of metastable
states. To develop this picture into a predictive dynamical theory I discuss
how to compute the exponentially small probability of a jump from one
metastable state to another. This is expressed as a path integral that can be
evaluated by saddle-point methods in mean-field models, leading to a boundary
value problem. The resulting dynamical equations are solved numerically by
means of a Newton-Krylov algorithm in the paradigmatic spherical $p$-spin glass
model that is invoked in diverse contexts from supercooled liquids to
machine-learning algorithms. I discuss the solutions in the asymptotic regime
of large times and the physical implications on the nature of the
ergodicity-restoring processes.
\end{abstract}

\maketitle

\section{Introduction}

The emergence of a rugged free energy landscape with many minima and saddles is a paradigm often invoked to explain complex dynamical phenomena ranging from supercooled liquids \cite{charbonneau2014fractal} to the performance of widely used algorithm in machine-learning and inference \cite{ros2019complex,mannelli2020marvels}.
Powerful methods developed originally in the spin-glass literature \cite{mezard1987spin} allow to obtain a rather complete description of the landscape in a growing number of statistical physics model. On the debit side  the resulting picture is essentially limited to {\it static} properties of the landscape, like the energy of the metastable states, their free energy and notably their number, the so-called configurational entropy.
Thus understanding the exact way in which  the landscape shapes the dynamics is largely an open problem.
The question is particularly urgent in the context of supercooled liquids where the rugged landscape paradigm is at the core of the Random-First-Order-Transition (RFOT) theory  \cite{kirkpatrick1989scaling,wolynes2012structural}.

More than thirty years after its formulation RFOT is still one of the major competing theories in the ongoing debate on the nature of the Glass transition \cite{rizzo2020critical}.
In a nutshell the theory posits that 
the physics of supercooled liquids is the same of Spin-Glass (SG) models displaying one-step of Parisi's Replica-Symmetry-Breaking (1RSB) \cite{mezard1987spin}. The mean-field versions of these models display an ergodicity-breaking transition at a dynamical temperature $T_d$ where the phase space splits into many metastable states that trap the dynamics; at lower temperatures, the configurational entropy, {\it i.e.} the log of the number of metastable states, decreases eventually vanishing at a static temperature $T_s$.
Ergodicity breaking between $T_d$ and $T_s$ is a mean-field artifact and one expects that in real systems ergodicity is restored through droplet-like excitations. Furthermore the size of these excitations must diverge as the configurational entropy vanishes  leading eventually to a genuine ergodicity-breaking transition at $T_s$.
RFOT originated from the realization \cite{kirkpatrick1987p} that similar features had been discussed in various unrelated (and themselves controversial) theories of supercooled liquids, most notably: 1) dynamics at the ergodicity-breaking transition is the same of the (avoided) Mode-Coupling-Theory (MCT) of supercooled liquids \cite{gotze2008complex}, 2) within the Adams-Gibbs-Di Marzio theory \cite{gibbs1958nature,adam1965temperature}, the glass transition is driven by a correlation length that diverges at the Kauzmann temperature where the configurational entropy vanishes.

Efforts to validate the theory have been driving theoretical, numerical and experimental research for years.
At the theoretical level RFOT has been substantiated by a number of results arguing that mean-field models of supercooled liquids exhibit 1RSB \cite{Mezard2012glasses,monasson1995structural}  including the solution of  the limit of infinite physical dimensions \cite{charbonneau2014fractal,charbonneau2017glass}. Numerically, the MCT phenomenology is well documented \cite{kob1995testing,kob1999computer} as well as the increase of dynamic \cite{kob1997dynamical,flenner2014universal} and static correlation lengths \cite{berthier2005direct,biroli2008thermodynamic}. Experimentally, the observation of a decreasing configurational entropy dates back to the 40's while more recently RFOT has inspired measurements of non-linear susceptibilities \cite{albert2016fifth}. 
Furthermore the analogy with supercooled liquids has also led to the discovery that off-equilibrium relaxational dynamics of mean-field Spin-glass models display aging \cite{cugliandolo1993analytical} and it is an active line of research  \cite{folena2020rethinking,Altieri_2020}.

In spite of this huge body of work, consensus on the validity of the theory is still lacking. One of the problem is that  RFOT-inspired literature often focuses on quantities, namely point-to-set correlation lengths and configurational entropy, whose actual relevance for the problem of the glass transition, that instead is dynamical in its essence, can also be questioned.
Besides, predictions are often merely qualitatively, which is a problem given that {\it e.g.} the observed static length increases are too modest to convince the community that they actually drive the slowing down of the dynamics.
\begin{figure}
\centering
\includegraphics[scale=.45]{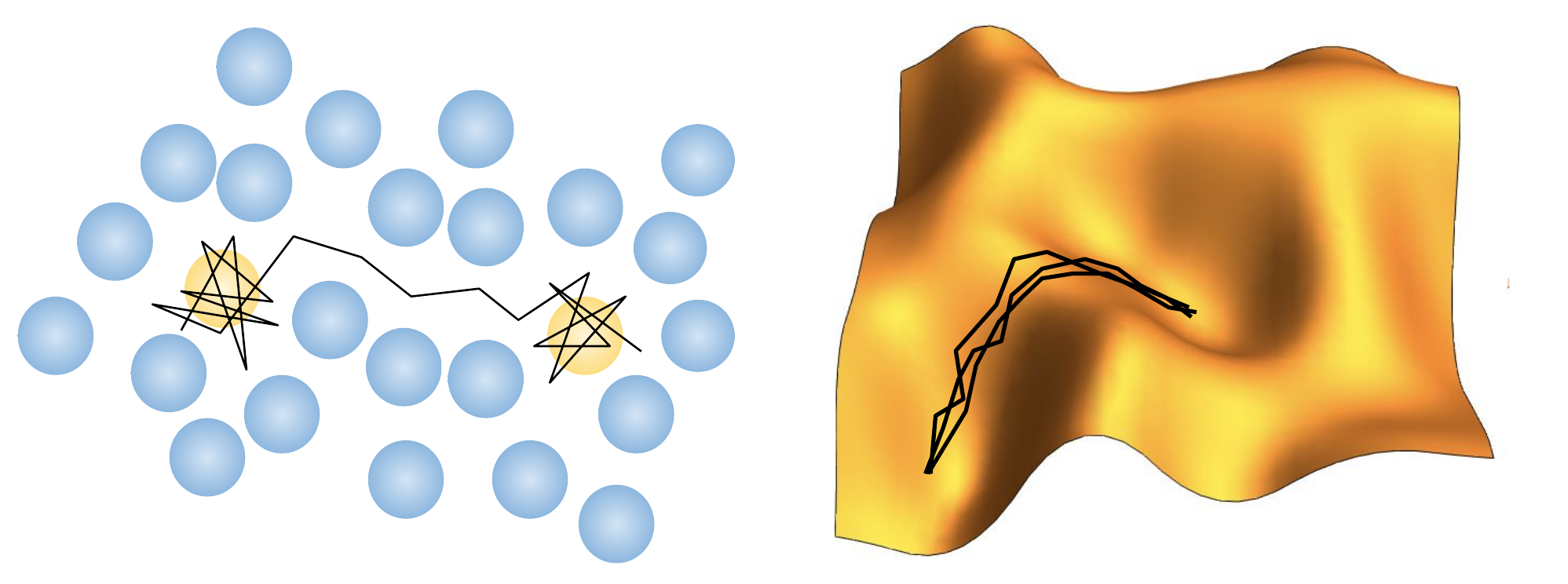} 
 \caption{Supercooled liquids display caging and hopping in real space (left) that correspond to evolution in the rugged phase space of mean-field models (right).}
\label{fig:cage-land}
\end{figure}
To make progress it would be important to obtain RFOT predictions that are  both {\it quantitative} and {\it dynamical}.
The main challenge is that current theoretical knowledge is mostly limited to mean-field models that display ergodicity-breaking at $T_d$: in order to recover ergodicity between $T_d$ and $T_s$ and obtain realistic predictions we have to go beyond mean-field, intense efforts in this line of research are currently underway \cite{baity2018activated,baity2018activatedb,
carbone2020effective,hartarsky2019maximum,
ros2019complexity,stariolo2019activated,
stariolo2020barriers,ros2021dynamical}.
In recent years progress  has been made for the temperature region {\it close} to the dynamical temperature $T_d$ \cite{rizzo2014long,rizzo2015qualitative,rizzo2016dynamical}. It is now possible to describe qualitatively and quantitatively how the ergodicity-breaking MCT transition is turned into a dynamical crossover in mean-field SG \cite{rizzo2016glass} (due to finite-size effects) and most importantly in some finite-dimensional models \cite{rizzo2020solvable}.
In this paper I consider instead the region between $T_d$ and $T_s$ where metastable states are present and discuss how to compute the transition rate,  that is {\it the exponentially small  probability} of a  jump from  an equilibrium state to another occurring in a finite time.
To make contact with the phenomenology of supercooled liquids we have to remember that below the experimental MCT transition temperature a particle is trapped most of the time into a cage formed by the surrounding particles and diffusion occurs through {\it hopping} {\it i.e.} sudden rare jumps from a cage to  another (see fig. \ref{fig:cage-land}).
Mean-field models capture {\it caging} through the appearance of metastable states and the study of jumps in the free energy landscape initiated in this paper is essential to a quantitative description of  {\it hopping} in real space.

\subsection{Main Results} 
 
The main results to be discussed in the paper are: i) a path integral method to compute the transition rate  ii) a Newton-Krylov algorithm yielding the numerical solution of the corresponding dynamical equations iii) an asymptotic analysis of the solutions in the regime of large times and iv) some non-trivial implications on the ergodicity-restoring processes.

\begin{figure}
\centering
\includegraphics[scale=.5]{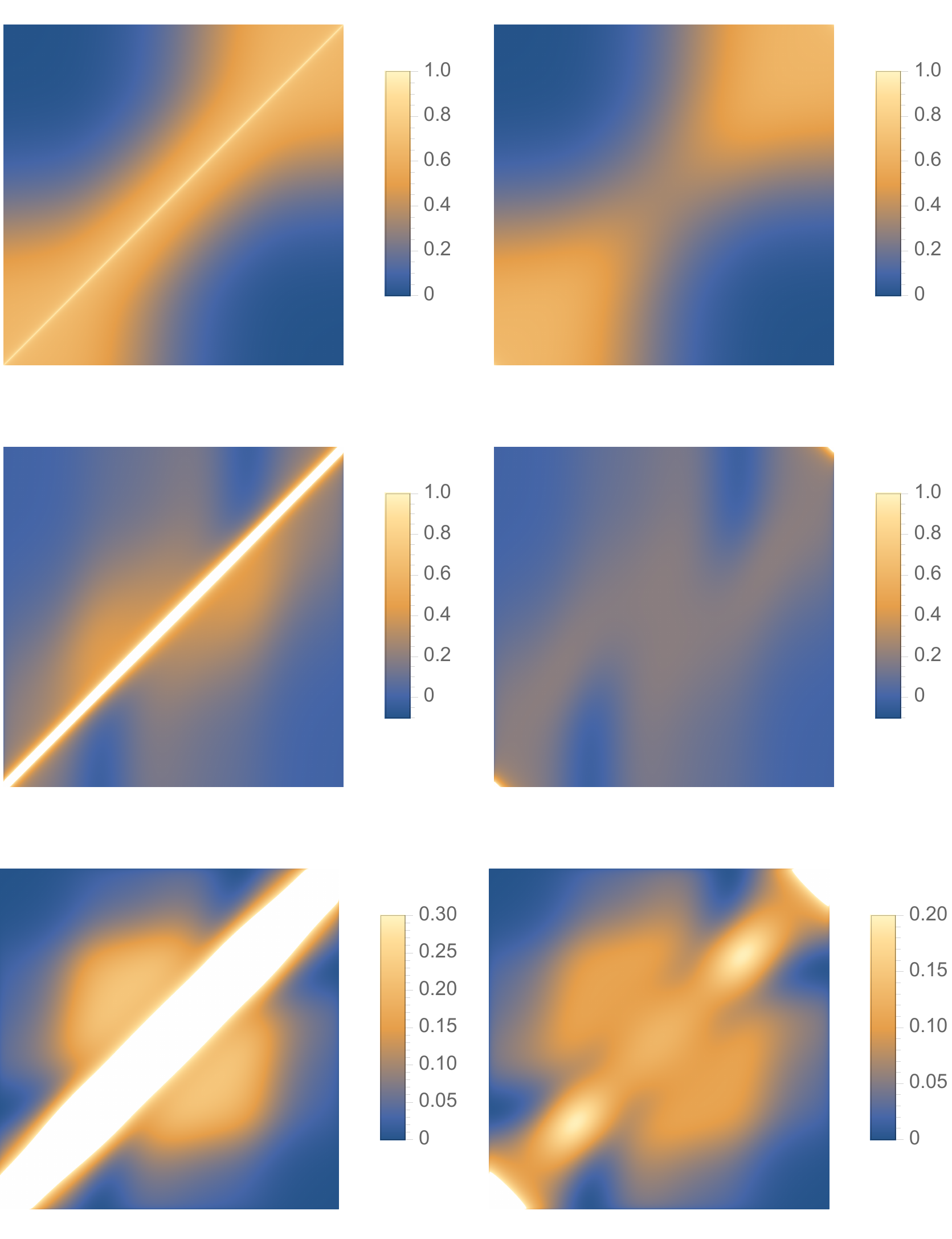} 
 \caption{left to right, top to bottom: density plot of the six functions $C(t,t')$, $C^{dt}(t,t')$, $\Tau \hR(t,t')$, $\Tau \hR^{dt}(t,t')$, $\Tau^2 \hX(t,t')$,  $\Tau^2 \hX^{dt}(t,t')$ for $-\Tau \leq t,t'\leq \Tau$ and $\Tau=40$ for the Spherical 3-SG model at $T=1/1.695<T_d$, (numerical solution with $\Delta t=\Tau/2000$). Too large values close the diagonal are not shown for clarity (white).}
\label{fig:sol}
\end{figure}
At the methodological level the problem is formulated as a path integral over Langevin dynamic trajectories that can be computed  though saddle-point methods in mean-field models.
The problem displays some important differences with respect to the standard relaxational dynamics \cite{sompolinsky1982relaxational,crisanti1993sphericalp,cugliandolo1993analytical}, namely the use of replicas and the need to explicitly handle the divergent path integral.
I have focused on the paradigmatic spherical $p$-spin SG model \cite{crisanti1992sphericalp,crisanti1993sphericalp} but the method is fairly general  and the equations can be derived with some effort for other mean-field systems {\it e.g.} supercooled liquids in large dimension  \cite{kurchan2016statics,maimbourg2016solution,manacorda2020numerical}.
Another more important difference  follows from to the fact that while in ordinary relaxation dynamics one only fixes the {\it initial} condition, in order to study activated dynamics one has to fix both the {\it initial} and {\it final} conditions.
This difference manifests itself at the level of the dynamical equations: while relaxational equations display first-order time derivatives the equations obtained here are second-order. One must then solve a more difficult boundary value problem instead of a simpler initial value problem \footnote{This is also the main difference with the earlier method of \cite{lopatin1999instantons,lopatin2000barriers}  
and more recently of \cite{ros2021dynamical}
that allow to study some special activated processes through relaxational dynamics equations but cannot be used to study transition rates between generic equilibrium states.}.
Indeed while the relaxational dynamical equations can be solved at times $t+\Delta t$ iteratively using the solution at times $t'\leq t$
the activated dynamical equations must be solved simultaneously at all times, something for which no standard algorithm exists.
A successful solution strategy has been developed 
 based on three elements: 1) Newton's method on discretized equations, 2) Krylov methods with physical preconditioning to invert the Jacobian, 3) Richardson extrapolation to reach the continuum limit. 
The whole technology can be again exported to other problems, with possible computational complexity issues due to the specific dynamical order parameter.

The paper is organized as follows.
In the remaining of this section I will give a compact presentation of the main results, leaving the details for the body of the paper and the appendices.
In particular in section \ref{sec:therate}
the object I compute is introduced and its general properties discussed. 
Later on I discuss the asymptotic behavior of the solutions that is interesting for a number of physical and technical reasons.
From the physical point of view the most interesting outcome of the computation is associated to the ergodicity restoring processes as discussed in sec. \ref{sub:ergres}.
The actual dynamical equations are given later in section (\ref{sec:analysis}), their solutions will be studied in various regimes (free, ergodic and activated). The numerical solution is challenging and will be discussed in sec. \ref{sec:num}. In section \ref{sec:conclu} I will give some concluding perspectives.

\subsubsection{The Transition Rate}
\label{sec:therate}

\begin{figure*}
\centering
\includegraphics[scale=.35]{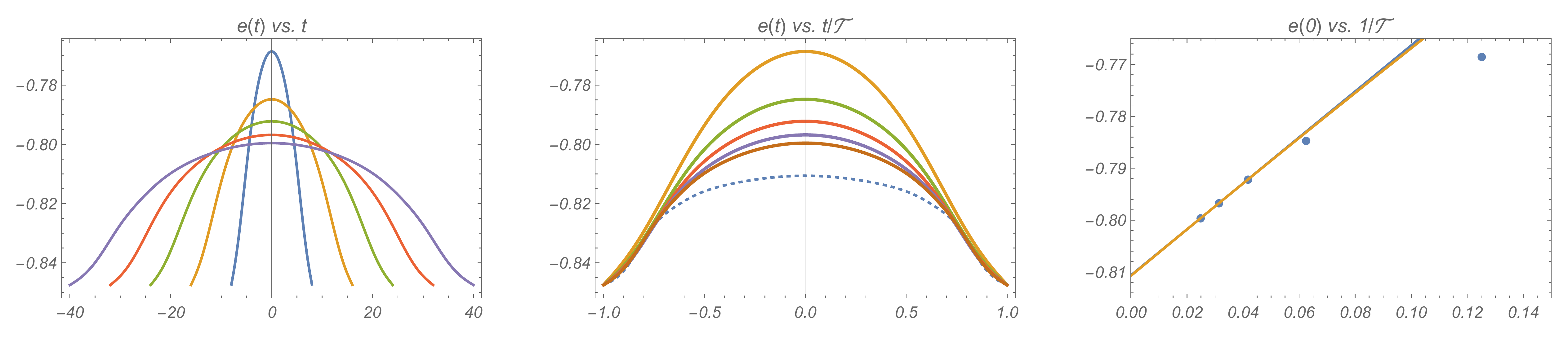}
 \caption{Spherical 3-SG at $T=1/1.695<T_d$. 
Left: instantaneous energy of the trajectory for   
  for $\Tau=8,\,16,\,24,\,32,\,40 $,  each curve was obtained from finite $N_t$ solutions through Richardson extrapolation to $\Delta t = 0$ (see text). Center: Same data with times rescaled with $\Tau$. The dotted line is the $\Tau=\infty$ limit obtained by linear extrapolation in $1/\Tau$.
Right: The energy at $t=0$ plotted vs. $1/\Tau$. Convergence to to the $\Tau=\infty$ limit with $1/\Tau$ corrections is clearly visible, the continuous lines are linear fits on the last two and three points. The energy varies continuously along the trajectory from the equilibrium  value $e_{eq}=-\beta/2=-0.8475$ at the initial and final time to larger values at intermediate times.  Note that the (extrapolated)  maximum value $e(0)=-0.8106$ at $t=0$ is larger than the threshold energy $e_{th}=-0.8311$, as computed from  eqs. (12,13) in \cite{cugliandolo1993analytical}  
. }
\label{fig:et-various}
\end{figure*}

The object considered is the transition rate $T_{2\Tau}(\sigma|\tau)$ defined as the probability that the system is in configuration $\sigma$ at time $t_{fin}=\Tau$ given that it was in configuration $\tau$ at time $t_{in}=-\Tau$. It is convenient to actually consider the following object that, due to detailed balance, is symmetric with respect to the exchange of $\sigma$ and $\tau$:
\beq
\hT_{2 \Tau}(\sigma,\tau) \equiv T_{2 \Tau}(\sigma|\tau)e^{{\beta \over 2} H(\sigma)-{\beta \over 2} H(\tau)}\ .
\eeq
An integral representation of Langevin dynamics is used and $\sigma$,$\tau$ are chosen as generic {\it equilibrium} configurations. At low temperatures  $\hT(\sigma,\tau)$ is exponentially small in the system size $N$ in mean-field models, therefore an annealed average $P_{eq}(\sigma)P_{eq}(\tau)\hT(\sigma,\tau)$ would interfere with the equilibrium measure of $\sigma$ and $\tau$ and
the correct procedure is to consider the quenched average
\beq
[\ln \hT] \equiv \sum_{\sigma,\tau} P_{eq}(\sigma)P_{eq}(\tau)\, \ln \hT(\sigma,\tau)\  \ .
\eeq
One can resort to the replica method to eliminate the logarithm, besides, if quenched disorder is present, the corresponding averages (represented by an overline in the following) require the introduction of  additional replicas of the initial and final configurations.

One can argue that the rate is self-averaging, meaning that most couples $(\sigma,\tau)$ display  the same rate \footnote{This can be shown computing $O(n)$ corrections to the rate where $n$ is the replica number.}, thus, given an initial configuration $\sigma$, the total number of configurations with rate equal to the typical value given by the average $\overline{[\ln \hT_{2 \Tau} ]}$ is equal to the total number of equilibrium configurations $\tau$, {\it i.e.} the exponential of the entropy, $e^S$.
Neglecting exponentially small corrections,  the total probability of jumping to one of the equilibrium configurations is thus $e^{[\ln \hT]+S}$ and must be smaller than one leading to the bound:
\beq
\overline{[\ln \hT_{2 \Tau} ]}+S \leq 0\ .
\label{bound}
\eeq
Now,  after a finite time a system in equilibrium will be in another equilibrium configuration {\it correlated} with the initial one, therefore the probability to be in a {\it generic} equilibrium configuration (that is uncorrelated to the initial one) must be smaller than one meaning that at any finite $\Tau$ the above bound should not be saturated. This also implies that even for $T>T_d$ if $\Tau<\infty$   one must consider the logarithm of the rate (and thus resort to the replica method with replica number $n=0$) in order not to interfere with the equilibrium measure on the initial and final conditions. On the other hand ergodicity implies that when $\Tau$ goes to infinity the probability measure will be flat over the $e^S$ equilibrium configurations and the average rate should become equal to $e^{-S}$ saturating the bound.

At the mean-field level the expression for the logarithm of the rate can be computed by the 
saddle-point approximation meaning that the 
thermodynamic limit $N \rightarrow \infty$ is always taken before the $\Tau \rightarrow \infty$ limit. Thus one can compute the quantity
\beq
\mathcal{A} \equiv \lim_{\Tau \rightarrow \infty} \lim_{N \rightarrow \infty}{1 \over N}\left(\overline{[\ln \hT_{2 \Tau} ]}+S \right)\ .
\label{defA}
\eeq 
In 1RSB models, as discussed in the introduction, the dynamical temperature $T_d$ marks the onset of activated dynamics and this results in  the two limits ceasing to commute:
\beqa
\mathcal{A}  & = &  0 \, \ \ \mathrm{for}\ T \geq T_d
\label{A0}
\\
\mathcal{A}  &  < &  0  \, \ \ \mathrm{for}\ T<T_d
\label{Am0}
\eeqa
In the following sections the expression for the average transition rate of the spherical $p$-spin model  for $T>T_s$ is given.
The Hamiltonian of the  model is given by:
\beq
H= \sum_{p=1}^{\infty} \sum_{i_1 < \dots < i_p} J_{i_1\dots i_p}s_{i_1}\dots s_{i_p}
\eeq
where the $N$ spins verify the global spherical constraint $\sum_i s_i=N$, the $J$'s are quenched Gaussian random variables of zero mean and variance:
\beq
\overline{J_{i_1\dots i_p}^2}={\mu_p \, p! \over 2 N^{p-1}} \, .
\eeq The following function is defined for convenience: 
\beq
f(x) \equiv \sum_{p=1}^\infty \mu_p \, x^p\ .
\eeq
The expression for the rate $\overline{[\ln \hT_{2 \Tau} ]}$ depends on
 six real functions $C(t,t')$, $\hR(t,t')$, $\hX(t,t')$, $C^{dt}(t,t')$, $\hR^{dt}(t,t')$ and $\hX^{dt}(t,t')$ defined for $-\Tau \leq t,t' \leq \Tau$. Two additional functions $\mu(t)$ and $\hat{\mu}(t)$ enforce the spherical constraint  leading to $C(t,t)=1$ and $\hR(t,t)=1/2$ for all $t$. Extremization of the expression leads to eight non-linear integro-differential equations (albeit two are redundant due to symmetries) that will be discussed in the next section.
The physical meaning of $C(t,t')$ and $C^{dt}(t,t')$ is straightforward:
\beqa
C(t,t') & = & \overline{[ \langle s_i(t) s_i(t') \rangle ]}
\\
C^{dt}(t,t') & = & \overline{[ \langle s_i(t)\rangle \langle s_i(t') \rangle ]}
\eeqa
where the  square brackets mean averages with respect to the dynamical trajectories at fixed initial and final configurations $(\sigma,\tau)$ and fixed disorder. Thus $C(t,t')$ is the average correlation between configurations visited by the same trajectory at times $t$ and $t'$ while $C^{dt}(t,t')$ is the correlation between configurations visited by {\it different trajectories} (hence the suffix {\it dt}). It follows that the equations must satisfy the boundary conditions $C(\pm\Tau,\pm\Tau)=C^{dt}(\pm\Tau,\pm\Tau)=1$ and $C(\pm\Tau,\mp\Tau)=C^{dt}(\pm\Tau,\mp\Tau)=0$.

In section \ref{sec:ergodic} the $T>T_d$ regime will be discussed, here one can show that in the $\Tau \rightarrow \infty$ limit the saddle-point equations admit a solution in which the six functions are expressed in terms of the equilibrium correlation $C_{eq}(0,t)$ and that the rate tends to $-S$ leading to $\mathcal{A}=0$.
This is possible because $C_{eq}(0,\infty)=0$ and the boundary condition $C(\pm\Tau,\mp\Tau)=0$ can be satisfied for $\Tau \rightarrow \infty$ by the equilibrium solution.
This is no longer true for $T < T_d$ because equilibrium dynamics is trapped and $C_{eq}(0,\infty) \neq 0$, as a consequence for $T<T_d$  the average logarithm of the rate is smaller than $-S$ also in $\Tau \rightarrow \infty$ limit meaning that $\mathcal{A}$ becomes negative {\it continuously} at $T_d$. 
In fig. \ref{fig:sol} the numerical solution for the spherical SG model with $p=3$ is shown for $T=1/1.695<T_d$  and $\Tau=40$. 
The transition from the {\it ergodic} $(T>T_d)$ to the {\it activated} ($T<T_d$) regime is marked by a {\it qualitative} change in the solutions. As it will be shown in section \ref{sec:ergodic} in the ergodic phase at large $\Tau$ the various functions approach equilibrium time-translational-invariant forms (see eqs. (\ref{eqC} ,\ref{eqR}, \ref{eqX})) and density plots like those of fig. (\ref{fig:sol}) tend to become symmetric with respect to the $t=t'$ axes. Instead in the activated phase the density plots for $C(t,t')$ and $C^{dt}(t,t')$ display the block structure visible in fig. (\ref{fig:sol}) at all values of $\Tau$, corresponding to the fact that the system close to the initial and final time performs essentially a relaxational dynamics in the corresponding states and jumps from one state to the other for $t, t' \approx 0$.  A quantity that also displays a qualitative change is the instantaneous intensive energy $e(t)$ of the system along the trajectory from one equilibrium configuration to another one. By definition we have that $e(\pm \Tau)$ is equal to the equilibrium value $e_{eq}$ at the corresponding temperature. In general $e(t)$ is larger than $e_{eq}$ at intermediate times but in the limit $\Tau \rightarrow \infty$ it tends to be equal to $e_{eq}$ at all times for $T>T_d$, while for $T<T_d$ $e(t)>e_{eq}$ also in the large $\Tau$ limit and indeed one finds that $e(t) \approx e_u(t/\Tau)$ where $e_u(x)$ is a universal function independent of $\Tau$ as shown by figure \ref{fig:et-various}. This kind of asymptotic behavior is also shared by the solution as we will further discuss in the following. 

Let us now mentions for completeness a few technical features of the problem that will be further discussed in the following sections.
At all temperatures, even in the ergodic phase $T>T_d$, the solutions cannot be expressed in terms of the equilibrium correlation if $\Tau$ is {\it finite}, however the equations can be solved analytically at finite $\Tau$ in the free case ($T=\infty$) in which the system performs a Brownian motion on the $N-1$ dimensional sphere (see section \ref{sec:free}). 
In general the expression for the average rate requires the computation of a path integral with an infinite normalization factor, a well-known pathology that is typically discussed in the context of the harmonic oscillator \cite{Zinn-Justin2002,Parisi1988}. In the non-interacting case one can use the harmonic oscillator formulas to derive the expression of the rate as a function of $\Tau$. 
Knowledge of the rate in the non-interacting case provides an alternative way to compute the rate at finite temperature without having to deal with the infinite normalization factor. Differentiating the saddle-point expression of the rate with respect to the inverse temperature $\beta$ one gets indeed a finite expression that yields the correct derivative when evaluated on the solutions of the saddle-point equations. The rate at finite $\Tau$ and $\beta$ can then be obtained by integration in $\beta$ starting from the exact $\beta=0$ result.

\subsubsection{Asymptotic Behavior}
\label{asymptotic}

The equations in the activated phase $T<T_d$ can only be studied numerically for finite $\Tau$. One then faces the problem of extrapolating the results to $\Tau \rightarrow \infty$ in order to avoid transient effects.
\begin{figure}
\centering
\includegraphics[scale=.27]{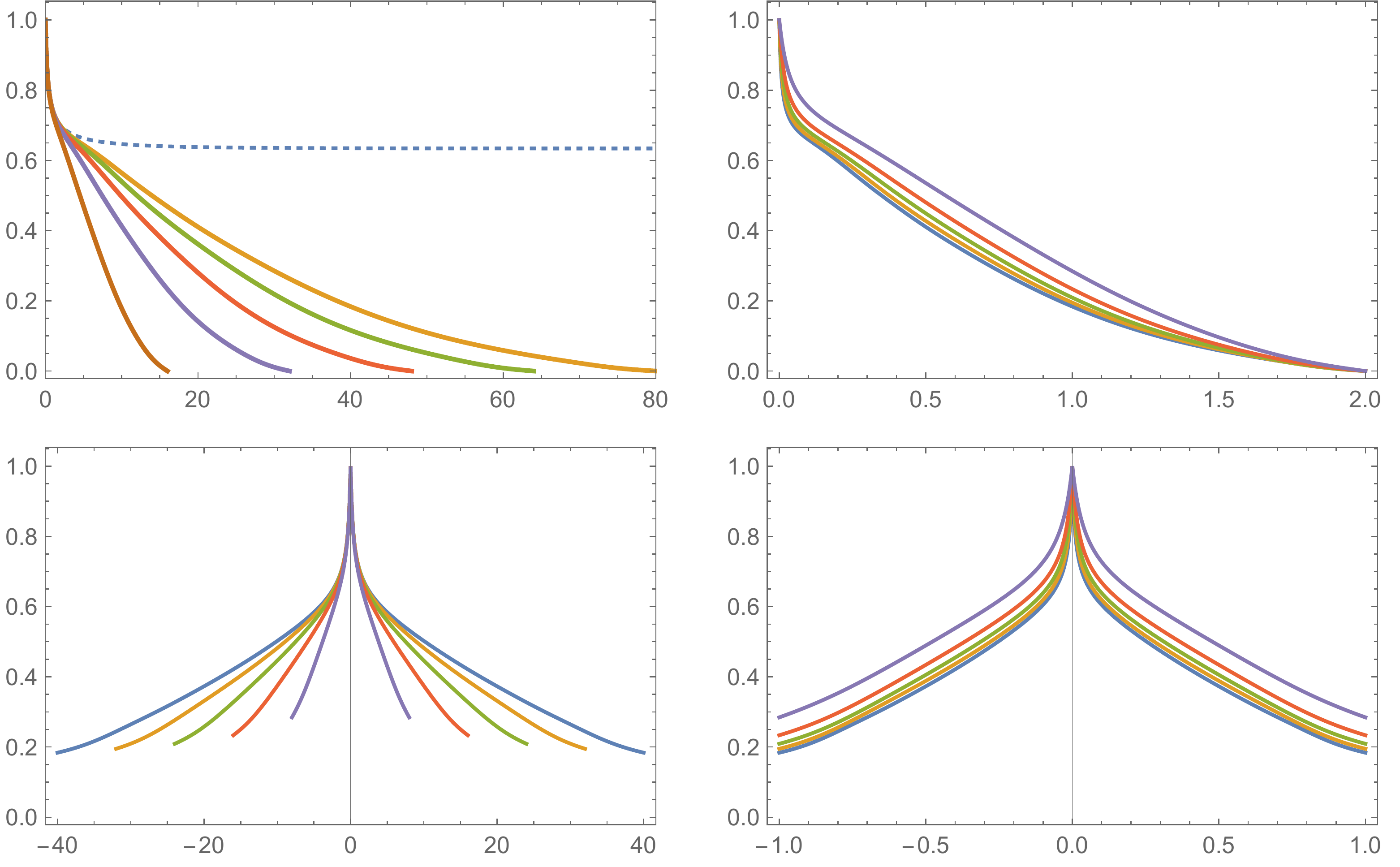} 
 \caption{Spherical 3-SG at $T=1/1.695<T_d$. Top, left: $C(-\Tau,t)$ vs. $t+\Tau$ for $\Tau=8,\,16,\,24,\,32,\,40 $, the dashed line is the equilibrium dynamics that is trapped inside a state and has a plateau at $0.6340$; right: same vs. $t/\Tau$. Bottom, left: $C(0,t)$ vs. $t$ for $\Tau=8,\,16,\,24,\,32,\,40 $, right: same vs. $t/\Tau$. Each curve was obtained from finite $N_t$ solutions through Richardson extrapolation to $\Delta t = 0$ (see text). }
\label{fig:sol-tau}
\end{figure}
The values of $\Tau$ that it was possible to reach numerically  allow to identify clearly the  {\it asymptotic behavior} for $\Tau \rightarrow \infty$ in the activated region. In particular in section (\ref{sec:ergodic})
we will show see that for finite time differences $|t-t'| =O(1)\ll \Tau$ the functions $C(t,t')$, $\hR(t,t')$ and $\hX(t,t')$ verify the equilibrium relationships corresponding to fluctuation-dissipation theorem and time-translational invariance meaning that on finite time-scales the trajectories are essentially equilibrium trajectories in a self-induced slowly-varying field.
Instead on the large time scales $|t-t'|=O(\Tau)$ the solutions approach universal functions independent of $\Tau$:
\beqa
C(t,t') & \approx & C_u(t/\Tau,t'/\Tau)
\\
C^{dt}(t,t') & \approx & C_u^{dt}(t/\Tau,t'/\Tau)
\\
\hR(t,t') & \approx & \Tau^{-1}\hR_u(t/\Tau,t'/\Tau)
\\
\hR^{dt}(t,t') & \approx & \Tau^{-1}\hR_u^{dt}(t/\Tau,t'/\Tau)
\\
\hX(t,t')& \approx & \Tau^{-2}\hX_u(t/\Tau,t'/\Tau)
\\
\hX^{dt}(t,t') & \approx & \Tau^{-2}\hX_u^{dt}(t/\Tau,t'/\Tau)
\\ 
\mu(t) & \approx & \mu_u(t/\Tau)
\\ 
\hat{\mu}(t) & \approx & \Tau^{-1}\hat{\mu}_u(t/\Tau)
\eeqa
The above behavior is clearly visible in fig. (\ref{fig:sol-tau}) for the function $C(t,t')$.
This also implies that the density plots for $C(t,t')$ and $C^{dt}(t,t')$ at different (large) value of $\Tau$ are indistinguishable.
Note also that $\hR(t,t')$ and $\hX(t,t')$ are small for $|t-t'|=O(\Tau)$ while they are finite close to the diagonal, see fig. \ref{fig:sol}. 

Plugging the above asymptotic expressions in the full dynamical equations one sees that 
 {\it the universal functions solve the dynamical equations with the second derivatives dropped}, similarly to what happens for equilibrium dynamics at $T_d$ and for off-equilibrium dynamics \cite{crisanti1993sphericalp,cugliandolo1993analytical} . Closing those equations would allow to work directly at $\Tau \rightarrow \infty $ and is an open problem that is left for future work.
At any rate inspection of the equation suggests that the solution should approach their   $\Tau=\infty$ limit with $1/\Tau$ corrections. This is indeed supported by the data as can be seen 
 considering {\it e.g.} the instantaneous (intensive) energy $e(t)$ along a trajectory, see the right panel of fig. (\ref{fig:et-various}).

One should note that the asymptotic structure of the solutions is utterly different from that of  metastability in ferromagnetism. The corresponding computation describes the transition rate from the metastable minimum to the stable one and leads to an instantonic equation in which the second order derivatives is not dropped in the asymptotic limit. As a consequence even if $\Tau \rightarrow \infty$ the jump effectively occurs in time window centered around $t=0$ that remains {\it finite}  and does not scale with $\Tau$.
Another more striking difference with metastability in ferromagnetism occurs when we consider the ergodicity-restoring processes as discussed next.

\subsubsection{Ergodicity Restoring Processes}
\label{sub:ergres}

This work is focused on the {\it exponentially small} probability that the system jumps to another equilibrium state in a {\it finite} time, which is complementary to the problem of determining the {\it exponentially large} time-scale $\tau_{erg}$ over which the system finds itself into another equilibrium state with {\it finite} probability.

Given that,  according to the discussion before eq. (\ref{bound}) the exponentially small probability to jump to an equilibrium state with typical rate is $p=e^{[\ln T]+S}$ it is natural to expect that such a probability becomes finite on a time-scale of order $1/p$. This is indeed what happens for ferromagnetism, {\it e.g.} in the Curie-Weiss model at low temperatures where two metastable states are present.
The following discussion shows instead that {\it the connection does not hold in presence of an exponential number of metastable states, {\it i.e.} a finite configurational entropy}. 
In order to see this it is convenient to introduce the average of the rate at inverse temperature $\beta$ over initial and final configurations that are in equilibrium at different inverse temperatures $\beta_{in}$ and $\beta_{fin}$:
\beq
[\ln T]_{\beta_{in},\beta,\beta_{fin}} \equiv \sum_{\sigma \, , \tau} P_{eq}^{(\beta_{fin})}(\sigma)P_{eq}^{(\beta_{in})}(\tau) \ln T^{(\beta)}_{2 \, \Tau}(\sigma|\tau) \ ,
\eeq
where $P_{eq}^{(\beta)}(\sigma)$ is the Boltzmann distribution at inverse temperature $\beta$ and $T^{(\beta)}_{2 \, \Tau}(\sigma|\tau)$ is the transition rate  from configuration $\tau$ to $\sigma$ due to Langevin dynamics at inverse temperature $\beta$.
The quantity that generalizes ${\cal A}$ as  given in eq. (\ref{defA}) is then:
\beq
{\cal A}(\beta'|\beta) \equiv \lim_{\Tau \rightarrow \infty} \lim_{N \rightarrow \infty}{1 \over N}\left(\overline{[\ln T_{2 \Tau} ]}_{\beta,\beta,\beta'}+S(\beta') \right)\ ,
\eeq
An expression of the above quantity can be easily obtained in terms of second-order dynamical equations as a straightforward generalization of the case where both the initial and final configurations are at equilibrium, see appendix \ref{sub:der}. 
One can thus compute the total probability of jumping  from an equilibrium state  to a state with a different energy $E' \neq E_{eq}$, as 
\beq
p(E') \propto \exp N {\cal A}(\beta'|\beta)
\eeq 
where naturally $\beta'$ is such that $E'$ is the equilibrium energy at the (inverse) temperature $\beta'$.
An explicit computation (see appendix \ref{sub:der}) shows that 
\beqa
\left.\frac{d p(E')}{d E'}\right|_{E'=E_{eq}}  & = &  0 \, \ \ \mathrm{for}\ T \geq T_d
\nonumber
\\
\left.\frac{d p(E')}{d E'}\right|_{E'=E_{eq}}  &  > &  0  \, \ \ \mathrm{for}\ T<T_d
\nonumber
\eeqa
The behavior for $T>T_d$ implies that the probability $p(E')$ has a maximum for $E'=E_{eq}$, this was to be expected due to the fact that $p(E_{eq})=1$ (see eq. (\ref{A0})) saturates the bound $p(E') \leq 1$.
Instead for $T<T_d$ we have $p(E') \gg p$ for $E' \geq E_{eq}$ and the system has a finite probability to jump to one of these higher energy states on a time scale $1/p(E')$ exponentially smaller than $1/p(E_{eq})$:
\beq
1/p(E') \ll 1/p(E_{eq}) \, \ \ \mathrm{for}\ T<T_d, E'>E_{eq}
\eeq
On the other hand the transition rate from a configuration with energy $E'$ back to a configuration with energy $E$ obeys the detailed balance condition:
\beq
e^{-\beta H(\tau)} T_{\Delta t}(\tau'|\tau)=e^{-\beta H(\tau')} T_{\Delta t}(\tau|\tau')\ ,
\eeq
taking into account that there are $S$ configurations with energy $E$ and $S'$ configurations with energy $E'$
we find that the probability to jump from a state with energy $E'$ {\it back} to an equilibrium state is
\beq
p(E')e^{\Delta S-\beta \Delta E} \gg p(E') \, ,
\eeq  
{\it i.e.} it is exponentially larger than $p(E')$ given that, by definition, the free energy $-S+\beta E$ has its minimum on the equilibrium states. Therefore, after the system has jumped to a state with higher energy $E'>E_{eq}$  it will  jump {\it back} to a {\it generic} equilibrium state on a scale exponentially smaller than $1/p(E')$, {\it i.e.} instantaneously on that scale. {\it This implies that an intermediate jump to one of the exponentially many metastable states with $E'>E_{eq}$ provides a more efficient path for restoring ergodicity than a direct jump to another equilibrium state and thus the  ergodic scale is smaller than $1/p(E_{eq})$, at variance with ferromagnetism where there are only two metastable states.} 
It is to be expected that $p(E')$ reaches a maximum $p(E_{max}) \ll 1$ at some $E_{max}>E_{eq}$ for $T<T_d$. The true ergodic time $\tau_{erg}$ should be identified with $1/p(E_{max})$ but could be even smaller, a detailed analysis of $p(E')$ and of the correct $\tau_{erg}$ is left for future work.

\subsubsection{Reproducibility}

The full commented code to solve numerically the equations is provided  online together with the required initialization files, see the ancillary files section of \cite{rizzo2020path}.

\section{Analysis of The Saddle-Point equations of the Spherical $p$-Spin-Glass Model}
\label{sec:analysis}

\subsection{The order parameter and its meaning}
\label{subsec:theor}

In the  appendix I will derive the expression for the logarithm of the rate $\mathcal{A}$ associated to Langevin dynamics of the Spherical $p$-SG model in terms on an order parameter determined through saddle-point equations.
The order parameter is a couple of  $2 \times 2$ matrices of functions of two times $t$ and $t'$ on the square $-\Tau \leq t,t' \leq \Tau $:
\beq
C \equiv \left(  \begin{array}{cc}
C(t,t') & \hR_2(t,t')
\\
\hR_1(t,t') & \hX(t,t')
\end{array}  \right)\, 
\label{defC}
\eeq
\beq
C^{dt} \equiv \left(  \begin{array}{cc}
C^{dt}(t,t') & \hR_2^{dt}(t,t')
\\
\hR_1^{dt}(t,t') & \hX^{dt}(t,t')
\end{array}  \right)\, 
\label{defCdt}
\eeq
The physical meaning of $C(t,t')$ and $C^{dt}(t,t')$ has been discussed already: $C(t,t')$ is the correlation on the same trajectory  (therefore $C(t,t)=1$ at all times in the spherical and Ising model) while $C^{dt}(t,t')$ measures the correlations between the configurations visited by {\it different trajectories}. Since by definition all trajectories have the same initial and final condition we have
\beq
C^{dt}(\pm \Tau,t)  =  C(\pm \Tau,t)\ .
\eeq
The correlations are obviously symmetric with respect to $(t,t')\rightarrow (t',t)$:
\beqa
C(t,t') & = & C(t',t)
\\
C^{dt}(t,t') & = & C^{dt}(t',t)\ .
\eeqa
Furthermore given that the measure over the trajectories is invariant under time reversal we have an additional symmetry with respect to the exchange of the initial and final configuration. Given that $t_{in}=-t_{fin}$ this symmetry translates  into:
\beqa
C(t,t') & = & C(-t,-t')
\\
C^{dt}(t,t') & = & C^{dt}(-t,-t')
\eeqa
For the $\hR$ components of the order parameter we have:
\beqa
\hR_1(t,t') & = & \overline{[\langle \hx_i(t)s_i(t')\rangle]}
\\
\hR_1^{dt}(t,t') & = & \overline{[\langle \hx_i(t) \rangle \langle s_i(t')\rangle]}
\eeqa 
where $\hat{x}$ is an auxiliary variable of the dynamics (see the appendix). They translate into:
\beqa
\hR_1(t,t') & = & \overline{ \left[{\delta\langle s_i(t')\rangle \over \beta\, \delta h(t)}\right]}-{1 \over 2}{d \over dt} C(t,t')
\\
\hR_1^{dt}(t,t') & = & \overline{ \left[{\delta \ln D(\sigma,\tau) \over  \beta\, \delta h(t)} \langle s_i(t')\rangle \right]}-{1 \over 2}{d \over dt} C^{dt}(t,t')
\eeqa 
Thus $\hR_1(t,t')$ is connected to the response of the time average of the spin over trajectories to a field $h(t)$. The functions $\hR_2(t,t')$ and $\hR_2^{dt}(t,t')$ are equal to the l.h.s.'s of the above equations with the exchange $t \leftrightarrow t'$. Thus while {\it neither} function is symmetric with respect to $t \leftrightarrow t'$ they are related through two functions $\hR(t,t')$ and $\hR^{dt}(t,t')$ such that
\beq
\hR_1(t,t') = \hR_2(t',t) = \hR(t',t) 
\eeq
\beq
\hR_1^{dt}(t,t') = \hR_2^{dt}(t',t) = \hR^{dt}(t',t) 
\eeq
which implies that the matrices (\ref{defC}) and (\ref{defCdt}) are symmetric.
On the other hand time-reversal invariance implies that $\hR_1(t,t')$,  $\hR_2(t,t')$,  $\hR_1^{dt}(t,t')$,  $\hR_2^{dt}(t,t')$, are symmetric with respect to $(t,t')\rightarrow (-t,-t')$ (because $t_{in}=-t_{fin}$ as above).
For the $\hX$ components we have:
\beqa
\hX(t,t') & = & \overline{[\langle \hx_i(t) \hx_i(t')\rangle]}
\\
\hX^{dt}(t,t') & = & \overline{[\langle \hx_i(t) \rangle \langle \hx_i(t')\rangle]}
\eeqa 
The physical meaning is also associated to particular responses that takes a simple form in the ergodic phase, see section \ref{sec:ergodic}. The above formulas imply that both $\hX(t,t') $ and $\hX^{dt}(t,t')$ are symmetric with respect to $t \leftrightarrow t'$ and to time-reversal $(t,t') \rightarrow (-t,-t')$.

\subsection{The Saddle Point Equations}
\label{sub:compact}

The saddle-point equations can be written in a compact form  which is also suitable for numerical integration  considering  the space of $2 \times 2$ matrices whose components are functions of two times $t$ and $t'$ on the square $-\Tau \leq t, t' \leq \Tau $. 
The generic element of this space can be written as 
\beq
A \equiv \left(  \begin{array}{cc}
C_A(t,t')  &  \hat{R}_{2,A}(t,t')
\\
\hat{R}_{1,A}(t,t') & \hX_A(t,t')
\end{array}  \right)\, 
\label{Agen}
\eeq
Given two elements $A$ and $B$ in the above space we have a natural definition of the product that generalizes the matrix product (it corresponds to exactly to ordinary matrix products if times are discretized).
For a real function $B(x)$ we also define the element-wise function $B[A]$ 
such that
\beqa
C_{B[A]}(t,t') & \equiv & B(C_A(t,t')) \, ,
\nonumber
\\
\hat{R}_{2,B[A]}(t,t') & \equiv & B'(C_A(t,t'))\, \hat{R}_{2,A}(t,t') \, ,
\nonumber
\\
 \hat{R}_{1,B[A]}(t,t') & \equiv & B'(C_A(t,t'))\, \hat{R}_{1,A}(t,t') \, ,
\nonumber
\\
\hX_{B[A]} (t,t') & \equiv  & B'(C_A(t,t'))\, \hX_A (t,t')+ 
\nonumber
\\ 
 & + & B''(C_A(t,t'))\, \hR_{1,A} (t,t')\hR_{2,A} (t,t') \, .
\nonumber
\eeqa
We define also: 
\beq
M   \equiv \left(  \begin{array}{cc}
 -{1 \over 2 \Gamma_0} \delta'' (t,t')  + \hat{\mu}(t)\delta(t,t')& \mu(t)\delta(t,t')
\\
\mu(t)\delta(t,t') &  -{2 \over \Gamma_0} \delta (t,t') 
\end{array}  \right)\, 
\label{Mope}
\eeq
and
\beqd
T \equiv \left(  \begin{array}{cc}
0 & \delta(t,t')
\\
\delta(t,t') & o
\end{array}  \right)\,  \ .
\eeqd
Note that due to the spherical constraint the operator $M$ depends on two  additional quantities $\mu(t)$ and $\hat{\mu}(t)$, see appendix \ref{sec:pimulti}.  
In order to write down the saddle-point equations it is useful to introduce two additional objects  $\Lambda$ and $\Lambda^{dt}$  that are also $2 \times 2$ matrices of two-time functions.
Another useful quantity is:
\beq
\delta_{\mp}   \equiv \left(  \begin{array}{cc}
 \hR_{\Lambda,1}(t,\mp\Tau)C(\mp\Tau,t')&\hR_{\Lambda,1}(t,\mp\Tau)\, \hR_2 (\mp\Tau,t')
\\
 C_{\Lambda}(t,\mp\Tau)C(\mp\Tau,t') &   C_{\Lambda}(t,\mp\Tau)\hR_2(\mp\Tau,t')
\end{array}  \right)\,  \ .
\eeq
With the above definitions the saddle-point equations of the spherical model derived in the appendix take the following compact expressions:
\beq
\Lambda = -{\beta^2 \over 2}f'[C]\, , \ \Lambda^{dt} = -{\beta^2 \over 2}f'[C^{dt}]
\eeq
\beqa
 M C + \,T\, \Lambda \, T \,  C  & + &  
\nonumber
\\
 +(n-1)  \,T\, \Lambda^{dt} \, T \,\, C^{dt}+  \delta_-+\delta_+  & = &   I
\label{spequa}
\\
M C^{dt} +\, T\, \Lambda^{dt} \, T\, \, C
 +  \, T\,\Lambda\, T\,  C^{dt} & + & 
\nonumber
\\
+ (n-2) \, T\, \Lambda^{dt} \, T\, C^{dt}+ \delta_-+\delta_+ & = &   0
\label{spequadt}
\eeqa
The quantities $\mu(t)$ and $\hat{\mu}(t)$ are unknown and must be determined self-consistently imposing the conditions:
\beq
C(t,t)=1 \,, \ \  \hR(t,t)={1 \over 2}  \ \ \ \forall t \ .
\label{condmu}
\eeq
Due to the presence of the operator $M$ that contains second-order derivatives the equations must be also supplemented with the following {\it boundary conditions}:
\beqa
C(\pm\Tau,t') & = & C(t',\pm \Tau)
\\
\hR_2(\pm \Tau,t') & = & \hR_1(t',\pm \Tau)
\\
C^{dt}(\pm \Tau,t') & = & C^{dt}(t', \pm \Tau)
\\
\hR_2^{dt}(\pm\Tau,t') & = & \hR^{dt}_1(t',\pm\Tau)
\eeqa 
Additional boundary conditions follow from the properties of the initial and final configurations:
\beqd
C(\pm \Tau,\pm \Tau)=1\, , \ \  C(\pm\Tau,\mp\Tau)=0
\eeqd
\beqd
C^{dt}(\pm \Tau,\pm \Tau)=1\, , \ \  C^{dt}(\pm \Tau,\mp \Tau)=0
\eeqd
The above boundary conditions are sufficient to compute the r.h.s. of the saddle point equations  (\ref{spequa},\ref{spequadt}) for a generic $C$ and $C^{dt}$ but given the structure of the equation and the physical meaning of the order parameter the actual solution verifies the additional symmetries discussed previously.
In particular the symmetry of the problem under the exchange between the initial and final configuration implies: 
\beq 
 \mu(t)=\mu(-t) \, , \ \   \hat{\mu}(t)=\hat{\mu}(-t)\ .
 \eeq
In the numerical analysis times are discretized, thus $C$ and $C^{dt}$ become actual matrices and the above symmetries allow a four-fold reduction of the memory required to store them.

\subsection{The free case}
\label{sec:free}

In the infinite temperature limit the interaction vanishes and the system performs a free Brownian motion on the $N-1$ dimensional sphere.
In this case one can find the analytic solution of the saddle-point equations which, in turn, is useful to guess an initial solution to feed to the Newton's algorithm at finite temperature.
Besides, as we discussed in the introduction, it allows to bypass the non-renormalizability of the expression for $\overline{[\ln \hT]}$ by integrating with respect to the temperature.
The saddle point equations (\ref{spequa},\ref{spequadt}) become in the free case:
\beqa
M C=I\,,\ \ M C^{dt}=0
\eeqa
The first equation corresponds to:
\beqa
-2 \hR_1(t,t')+\mu(t)C(t,t') & = 0
\nonumber
\\
-{1 \over 2}{d^2 \over dt^2} C(t,t')+ \mu(t)\hR_1(t,t')+\hat{\mu}(t)C(t,t') & = & \delta(t-t')
\nonumber
\\
-2 \hX(t,t')+\mu(t)\hR_2(t,t') & = & \delta(t-t')
\nonumber
\\
-{1 \over 2}{d^2 \over dt^2} \hR_2(t,t')+ \mu(t)\hX(t,t')+\hat{\mu}(t)\hR_2(t,t') & = & 0
\nonumber
\eeqa
Note that the equations for $C$ do not depend on $C^{dt}$. To solve them one can start noticing that the conditions (\ref{condmu}) plugged into the first equation lead to 
\beq
\mu(t)=1  \ \ \ \forall t \, .
\eeq 
Then the first equation and the symmetries $\hR_2(t,t')=\hR_1(t',t)$, $C(t,t')=C(t',t)$  and the third equation lead to
\beqa
\hR_1(t,t') & = & {1 \over 2} C (t,t')
\\
\hR_2(t,t') & = & {1 \over 2} C (t,t')
\\
\hX(t,t') & = & -{1 \over 2} \delta(t-t')+{1 \over 4}C(t,t') \ .
\eeqa
As a consequence the correlation $C(t,t')$ is determined by the following equation:
\beqd
-{1 \over 2}{d^2 \over dt^2} C(t,t')+ \left({1 \over 2}+\hat{\mu}(t)\right) C(t,t')  =  \delta(t-t')
\eeqd
To be solved with boundary conditions $C(\pm\Tau,t)=C(t,\pm\Tau)$ and $C(\pm \Tau,\pm \Tau)=1$, $C(\pm \Tau,\mp \Tau)=0$.
The equation has the solution
\beq
C(t,t')=C(|t-t'|) \,, \ \  \hat{\mu}(t)=\hat{\mu} \, .
\eeq
To determine the function $C(x)$ it is convenient to define a quantity $ -\infty < a < 1$  implicitly through the equation: 
\beq
 \Tau   =   {1 \over 2 \sqrt{a}} \,  \arctanh \sqrt{a} \ .
\eeq
\begin{figure*}
\centering
\includegraphics[scale=.8]{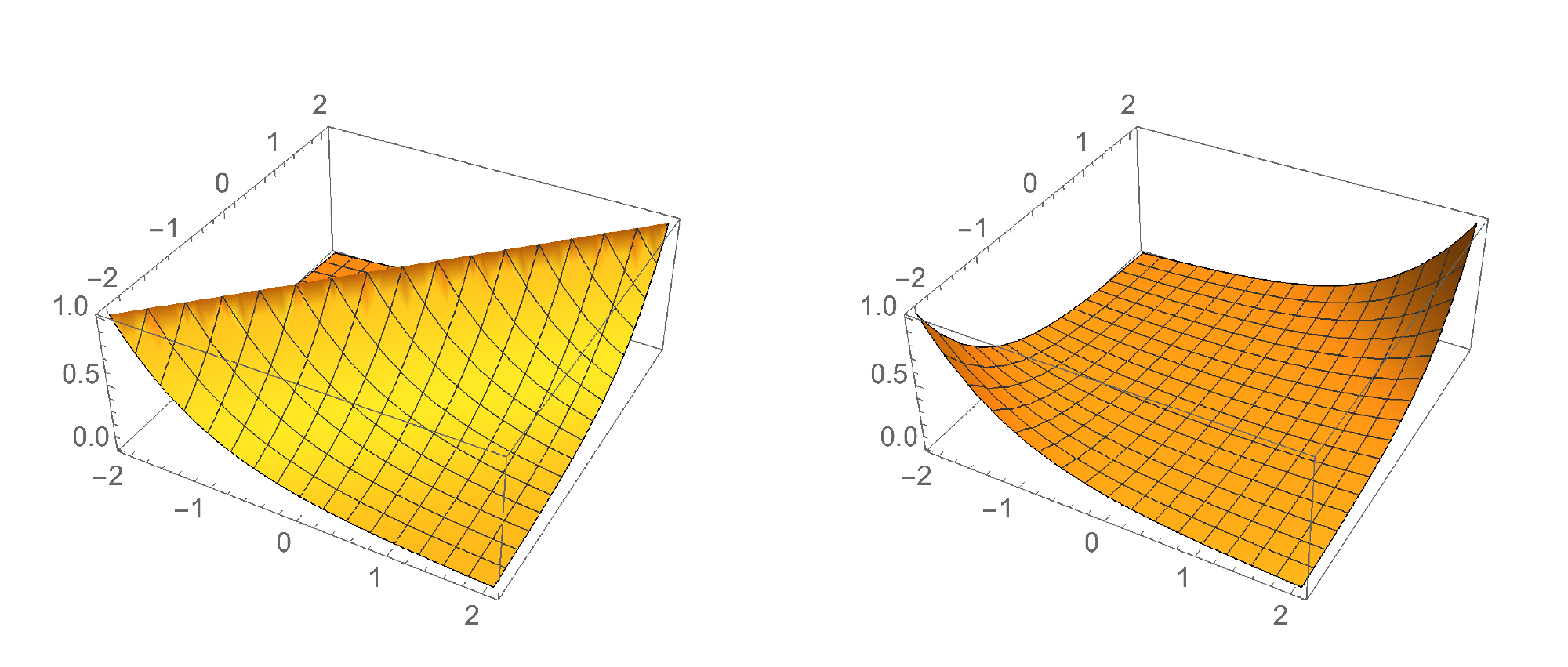} 
 \caption{Left: The function $C(t,t')$ in the free case for $a=0.999$ corresponding to $\Tau=2.07442$ (see text), Right: the function $C^{dt}(t,t')$.}
\label{fig:free}
\end{figure*}
We then have, see fig. (\ref{fig:free}): 
\beqa
\hat{\mu} &   = &    {a -1 \over 2}
\\
C(x) & = &  \cosh \sqrt{a} x - {1 \over \sqrt{a}} \sinh  \sqrt{a} x \ .
\label{freesol}
\eeqa
Note that we also have $C(0)=1$ and  $\dot{C}(0)=-1$  and these are general results  that hold also at finite temperature being a consequence of the presence of the delta functions and of the symmetry  $C(t,t')=C(t',t)$.
Let us discuss the limits of large and small $\Tau$. For large values of $\Tau$ we have:
\beq
 a \approx 1 - 4 \, e^{- 4 \Tau}\, ,\  \hat{\mu} \approx - 2 \, e^{-4 \Tau } \ \  \mathrm{for} \ \  \Tau \rightarrow \infty \ .
\eeq
Note that in the large $\Tau$ limit $\hat{\mu}$ tends to zero and  the function $C(x)$ tends to the equilibrium solution $
C_{eq}(x)= e^{-x}$.
On the other hand $a$ becomes negative for $\Tau<1/2$:
\beq
a \approx 6 \left(\Tau-{1 \over 2} \right )  \  \mathrm{for} \ \  \Tau \approx {1 \over 2} 
\eeq
therefore  $a^{1/2}$ is imaginary and the hyperbolic functions in (\ref{freesol}) become ordinary trigonometric functions:
\beqa
C(x) & = &  \cos \sqrt{|a|} x - {1 \over \sqrt{|a|}} \sin  \sqrt{|a|} x \ .
\eeqa
In the small $\Tau$ limit $a$ tends to minus infinity as 
\beq
a \approx - { \pi^2 \over 16 \Tau^2}  \, ,\  \hat{\mu} \approx  - { \pi^2 \over 32 \Tau^2} \ \  \mathrm{for} \ \  \Tau \rightarrow 0 
\eeq
leading to:
\beq
C(x) \approx \cos {\pi x \over 4 \Tau} \  \mathrm{for} \ \  \Tau \rightarrow 0 
\eeq
The above expression correctly vanishes at $x= 2 \Tau$ but does not display the correct behavior $C(x)\approx 1- x$ at small $x$. Indeed it is only valid for $x =O(\Tau)$ and the correct linear behavior is recovered for $x \ll \Tau$.
We now turn to the equations for $C^{dt}$:
\beqa
-2 \hR_1(t,t')+\mu(t)C^{dt}(t,t') & = &  0
\nonumber
\\
-{1 \over 2}{d^2 \over dt^2} C^{dt}(t,t')+ \mu(t)\hR_1^{dt}(t,t')+\hat{\mu}(t)C^{dt}(t,t') & = & 0
\nonumber
\\
-2 \hX^{dt}(t,t')+\mu(t)\hR_2^{dt}(t,t') & = & 0
\nonumber
\\
-{1 \over 2}{d^2 \over dt^2} \hR_2^{dt}(t,t')+ \mu(t)\hX^{dt}(t,t')+\hat{\mu}(t)\hR_2^{dt}(t,t') & = & 0
\nonumber
\eeqa
The condition $\mu(t)=1$ derived earlier, the first equation, the symmetries $\hR_2^{dt}(t,t')=\hR_1^{dt}(t',t)$, $C^{dt}(t,t')=C^{dt}(t',t)$  and the third equation lead to:
\beqa
\hR_1^{dt}(t,t') & = & {1 \over 2} \,C^{dt} (t,t')
\\
\hR_2^{dt}(t,t') & = & {1 \over 2}\, C^{dt} (t,t')
\\
\hX^{dt}(t,t') & = &  {1 \over 4}\,C^{dt}(t,t') 
\eeqa
and $C^{dt}(t,t')$ obeys the equation
\beq
-{1 \over 2}{d^2 \over dt^2} C^{dt}(t,t')+ \left({1 \over 2}+\hat{\mu}\right) C^{dt}(t,t')  =  0
\eeq
To be solved with boundary conditions $C^{dt}(\pm\Tau,t)=C^{dt}(t,\pm\Tau)$ and $C^{dt}(\pm \Tau,\pm \Tau)=1$, $C^{dt}(\pm \Tau,\mp \Tau)=0$.
The solution reads:
\beqa
C^{dt}(t,t') &  = &  {\sqrt{1-a} \over a}  \cosh (\sqrt{a} (t+t')) +
\nonumber
\\
& - &  {1-a \over a} \cosh(  \sqrt{a}  (t-t')) \, , \ \mathrm{for} \ \Tau>1/2
\nonumber
\\
C^{dt}(t,t') &  = &  {\sqrt{1-a} \over a}  \cos (\sqrt{|a|} (t+t'))+
\nonumber
\\
& - &  {1-a \over a} \cos(  \sqrt{|a|}  (t-t')) \, \ \mathrm{for} \ 0<\Tau<1/2
\nonumber
\eeqa
In the limit $\Tau \rightarrow \infty$ we have $C^{dt}(t,t')=0$ for any finite $t,t'$. Close to the initial and finite time we have instead
\beqd
C^{dt}(\pm \Tau+\delta t,\pm \Tau +\delta t')=C_{eq}(|\delta t+\delta t'|)\  \mathrm{for} \ \  \Tau \rightarrow \infty .
\eeqd
In the limit $\Tau \rightarrow 0$ we obtain instead:
\beqd
C^{dt}(t,t')\approx  \cos {\pi (t-t') \over 4 \Tau}  \  \mathrm{for} \ \  \Tau \rightarrow 0 
\eeqd
Note that in this limit we have also $C^{dt}(t,t') \approx C(t,t')$ meaning that all trajectories tend to follow the same path.

The computation of the transition rate requires to treat carefully the divergences of the path integral and is given in the appendix. The final result is:
\beqd
{1 \over N}\overline{[\ln \hT]}= -{1 \over 2}\ln 2 \pi-{1 \over 2} + (1+\hmu) \Tau +{1 \over 4} \ln {- \hmu \over 2}\ .
\eeqd 
From the above expression we see that the logarithm of the rate tends to minus infinity as $\Tau$ goes to zero:
\beqd
{1 \over N}\overline{[\ln \hT]} \approx -{\pi^2 \over 32 \Tau} \, ,\ \ \ \Tau\approx 0  \ .
\eeqd 
For $\Tau$ going to infinity one can see that although $\hat{\mu} \approx - 2 \, e^{-4 \Tau }$ the $O(\Tau)$ divergences in the  last two terms cancel and the rate has a finite limit. This limit is exactly equal to minus the entropy $S(\beta)$ of the spherical model at infinite temperature ($\beta=0$), {\it i.e.} the logarithm of the surface of the $N$-dimensional sphere of radius $\sqrt{N}$:
\beqd
\lim_{\Tau \rightarrow \infty }{1 \over N}\overline{[\ln \hT]}  = -{1 \over 2}\ln 2 \pi-{1 \over 2} =-S(0)\ .
\eeqd
This is expected  in the $\Tau \rightarrow \infty$ limit at any finite $N$ and implies that the two limits commute.
One can show that the approach to the  $\Tau \rightarrow \infty$ limit is exponential:
\beqd
{1 \over N}\overline{[\ln \hT]} \approx  -S(0)-{1 \over 2}e^{-4\, \Tau}\, , \ \ \Tau \gg 1 \ .
\eeqd

\subsection{The Ergodic Phase}
\label{sec:ergodic}

As explained in the introduction the ergodic phase defined by $T>T_d$ is characterized by the fact that the $\Tau \rightarrow \infty$ limit commutes with the $N \rightarrow \infty$ limit. 
In order to discuss the solutions in this regime it is convenient to analyze first the ergodic limit at finite $N$.
While the present formalism is fully invariant under time reversal   to discuss this limit it is convenient to reintroduce the arrow of time. In the ergodic limit $\Tau \rightarrow \infty$ dynamics looses any dependence on the initial configuration and the transition rate obeys
\beq
\lim_{\Tau \rightarrow \infty}T_{2 \Tau}(\sigma|\tau)={e^{-\beta H(\sigma)} \over Z}
\label{ergex}
\eeq
 that, according to the previous definitions leads to:
\beq
\lim_{\Tau \rightarrow \infty}\hT_{2 \Tau}(\sigma,\tau)={1 \over Z}e^{-{\beta \over 2} H(\sigma)-{\beta \over 2} H(\tau)} 
\eeq
and the following exact result:
\beq
\lim_{\Tau \rightarrow \infty}[\ln\hT_{2 \Tau}(\sigma,\tau)]=-\beta E-\ln Z=-S
\eeq
In the ergodic limit we expect that trajectories close to the initial condition at time $-\Tau$ are not influenced by the fact that we are fixing the final configuration at time $\Tau$. This implies that they are typical trajectories and since the initial configuration is weighted with the equilibrium weight we expect that correlations and response are those valid at equilibrium and in particular satisfy time-translational-invariance (TTI) and fluctuation-dissipation theorem (FDT).
To see the implications on the functions it is convenient to start from the expressions of the six functions  as averages of $s_i(t)$ and $\hat{x}_i(t)$ obtained in subsection \ref{subsec:theor}.
It is then convenient to transform back to the variable $\hat{s}$ introduced in the first steps of the integral represenation (see appendix \ref{sec:pathint}) by writing
\beq
\hat{x}_i= \hat{s}_i(t) - \dot{s}_i(t)\ .
\eeq
It is well known the equilibrium averages of $\hat{s}_i(t)$ are associated to responses to a field at time $t$ and since according to (\ref{ergex}) the distribution at $\sigma$ is completely independent of what happens at any finite time $t \ll \Tau$ we have for these times 
\beq
\langle \hat{s}(t) \rangle=0\, , \ \  \langle \hat{s}(t)\hat{s}(t') \rangle=0 \ .
\eeq
The above relationships plus FDT allow to derive the following expressions in the ergodic limit:
\beqa
C(t,t') & = & C_{eq}(|t-t'|)
\label{eqC}
\\
\hR_{2}(t,t') & = & \hR_{1}(t',t)= \sign(t-t') \, {1 \over 2}\,{d \over dt'}C(t,t')
\label{eqR}
\\
\hX(t,t') & = & -{1 \over 4}{d^2 \over dt\, dt'}C(t,t')
\label{eqX}
\eeqa
Note that $\hR(t,t')$ is symmetric (which is not true in the activated phase) and it is equal to $1/2$ on the diagonal according to the saddle-point equations.
The expression for $\hX(t,t')$ is valid also for $t=t'$ and thus it may be integrated, this is consistent with the fact that $\hX(t,t')$ has a term proportional  $-\delta(t-t')/2$ on the diagonal.
The above properties have been derived for $t,t' \ll \Tau$, in order to study the region of times close to $\Tau$ it is better to reverse the arrow of time, following the same arguments we obtain the above equations with $(t,t')\rightarrow (-t,-t')$ and since the functions are symmetric with respect to this transformation we conclude that they are valid at all times.

At large values of $\Tau$ the correlation between different trajectories are different from zero only  for times $t,t'$ that are both close to either $-\Tau$ or $\Tau$. To determine them we can use the following relationship :
\beq
\sum_{\tau}{e^{-\beta H(\tau)} \over Z}T_{\Delta t}(\tau'|\tau)T_{\Delta t'}(\tau''|\tau)={e^{-\beta H(\tau')} \over Z}T_{\Delta t+ \Delta t'}(\tau''|\tau')
\eeq
that follows from the detailed balance condition:
\beq
e^{-\beta H(\tau)} T_{\Delta t}(\tau'|\tau)=e^{-\beta H(\tau')} T_{\Delta t}(\tau|\tau')
\eeq
and from the general property:
\beq
\sum_{\tau}T_{\Delta t'}(\tau''|\tau)\,T_{\Delta t}(\tau|\tau')=T_{\Delta t+ \Delta t'}(\tau''|\tau')\ .
\eeq
Furthermore responses between different trajectories vanish because of  $\langle \hat{s}(t) \rangle=0$.
In particular if we write times as $ t =\mp \Tau \pm \Delta t$ ($\Delta t \geq 0$) we have:
\beqa
C^{dt}(t,t') & = & C_{eq}(\Delta t+\Delta t')
\label{eqCdt}
\\
\hR_{2}^{dt}(t,t') & = & \hR_{1}^{dt}(t',t)=  \mp {1 \over 2}\,{d \over dt'}C^{dt}(t,t')
\label{eqRdt}
\\
\hX^{dt}(t,t') & = & {1 \over 4}{d^2 \over dt\, dt'}C^{dt}(t,t')\ .
\label{eqXdt}
\eeqa
Note that the expression for $\hR_{2}^{dt}(t,t') $ changes sign depending on weather we are close to $\mp \Tau$, indeed the second case ($+\Tau$) is obtained from the transformation $(t,t')\rightarrow (-t,-t')$ applied to the first case ($-\Tau$).
A solution of the saddle-point equations with the above structure can be found only if $C_{eq}(\infty)=0$, thus the ergodic solution does not exist below $T_d$ where $C_{eq}(\infty)=q$ in the thermodynamic limit and the condition $C(\pm\Tau,\mp \Tau)=0$ cannot be fulfilled. 
The above expressions are valid for both Ising and spherical models, in addition for the spherical model we have:
\beq
\mu(t)= \mu_{eq}\, , \ \ \ \hmu(t)=0\, .
\label{mueq}
\eeq
Note that all components are parameterized by the equilibrium correlation $C_{eq}(t)$,  let us see how an equation for $C_{eq}(t)$ can be derived in the present context. In the formulation of the problem discussed in the appendix one works with objects $\bm{Q(ab)}$. In the ergodic phase one can additionally introduce a special class of objects $\bm{A(ab)}$ whose components $A(ab)$ and $A^{dt}(ab)$ obey formulas (\ref{eqC},\ref{eqR},\ref{eqX}) and (\ref{eqCdt},\ref{eqRdt},\ref{eqXdt}) with $C_{eq}(t)$ replaced by some generic function $C_{A}(t)$.
In the following this kind of objects will be loosely referred as having an `equilibrium''  structure.  A simple computation shows that the equilibrium structure is preserved by the application of a function, {\it i.e.} $f(\bm{A(ab)})$ has also the equilibrium structure with:
\beq
f(\bm{A(ab)}) \rightarrow C_{f[A]}(t)= f(C_{A}(t))
\eeq
Similarly, given another equilibrium object $\bm{B(ab)}$ parameterized by a function $C_B(t)$, one can show through a tedious computation that the product has also the equilibrium structure with:
\beqd
\int \bm{A(ac)B(cb)dc}  \rightarrow C_{AB}(t)
\eeqd
\beqd
 C_{AB}(t)=C_A(0)C_B(t) -  \int_0^t C_A(t-s) {d C_B(s) \over ds} ds
\eeqd
The above equation holds for $C_A(\infty)=C_B(\infty)=0$ which is granted by the fact that we are working in zero field and $f'(0)=0$ so that the overlap between different equilibrium states is zero. 
The last two equations applied to the 
saddle-point equation (\ref{fuleq}) 
lead immediately to the equilibrium equation: 
\beq
{\dot C}(t) = - \, C(t) -{\beta^2 \over 2}\int_0^t f'(C(t-s))\, \dot{C}(s)\, ds
\eeq
where we have used $\mu=1+{\beta^2 \over 2}f'(1)$ that follows from the condition $C(0)=-\dot{C}(0)=1$.
The above equations is usually written as \cite{crisanti1992sphericalp,crisanti1993sphericalp,castellani2005spin}:
\beqa
\dot{C}(t) & = &  - \, C(t) +{\beta^2 \over 2}f'(C(t))(1-C(t))+
\nonumber
\\
&-&{\beta^2 \over 2}\int_0^t (f'(C(t-s))-f'(C(t)))\, {dC \over ds} \, ds
\eeqa
Where $C(\infty)$ is the solution of the equation
\beq
C ={\beta^2 \over 2}f'(C)(1-C) \ .
\eeq
This equation admits the solution $C=0$ at all temperature but develops an additional non-zero solution at $T_d$ specified by the condition:
\beq
1 ={\beta_d^2 \over 2}(f''(C_d)(1-C_d)-f'(C_d))
\eeq
In the pure $p$-spin models $f(x)=x^p$ we have
\beq
T_d=\sqrt{{p (p-2)^{p-2} \over 2 (p-1)^{p-1}}}\,, \ \  C_d(\infty)={p-2 \over p-1}\ \ . 
\eeq

To conclude the discussion of the ergodic phase we have to show that for $T > T_d$ the $\Tau \rightarrow \infty$ and $N \rightarrow \infty$ limits commute and we have
\beq
\mathcal{A}=0 \, \ \ \mathrm{for}\ \,  T_d < T < \infty 
\label{aer}
\eeq
In the previous section we have shown  that the relationship is satisfied for $\beta=1/T=0$ and thus it suffices to show that its derivative with respect to the $\beta$ is also zero.
In appendix \ref{sub:der} we show that the derivative of the rate with respect to $\beta$ is $ {\beta \over 2 }{d \over d\,n}\int d{\bf a}d{\bf b}f({\bf Q(a,b)})$. The above expression takes a very simple form in the ergodic phase, indeed  using the above equilibrium formulas one can show that the generic  formula (\ref{intgen}) for the integrals takes a very simple form:
\begin{widetext}
\beqa
\int \bm{B(ab) da db}  & = &  (m+m')C_B(0) + n \left(   4 \int R_{2,B}(-\Tau,t) dt  + \int  \hX_B(t,t')\,dt \,dt'+(n-1)\int \hX_{B^{dt}}(t,t')\,dt\, dt' \right)= 
\nonumber
\\
& = & (m+m'+n)C_B(0) 
\label{inteq}
\eeqa
\end{widetext}
that leads to:
\beq
{d \over d\beta}\lim_{\Tau \rightarrow \infty }\lim_{N \rightarrow \infty}{1 \over N}\overline{[\ln \hT]} ={\beta \over 2}f(1)\, ,
\eeq
to be compared with
\beq
{1 \over N} {d S(\beta) \over d \beta} =-{\beta \over 2}f(1)\ .
\eeq
We thus see that eq. (\ref{aer}) is verified:
\beq
\lim_{\Tau \rightarrow \infty }\lim_{N \rightarrow \infty}{1 \over N}\overline{[\ln \hT]} = -S(\beta)=-S(0)+{\beta^2 \over 4}f(1)\ .
\eeq
The above relationship is valid also for Ising systems.

\subsection{The Activated Phase}

In the activated phase $T_s<T<T_d$ the saddle-point equations have been solved  numerically for the classic pure $p$-spin with $f(x)=x^3$ where $\beta_d=1.63299$ and $\beta_s=1.70633$ \cite{crisanti1992sphericalp,crisanti1993sphericalp,castellani2005spin}.
Due to the significant resources needed to solve the equations, as discussed in the next section,
most of the analysis has been done at a single inverse  temperature $\beta=1.695$ for values of $\Tau=8,16,24,32,40$.
\begin{figure*}
\centering
\includegraphics[scale=.35]{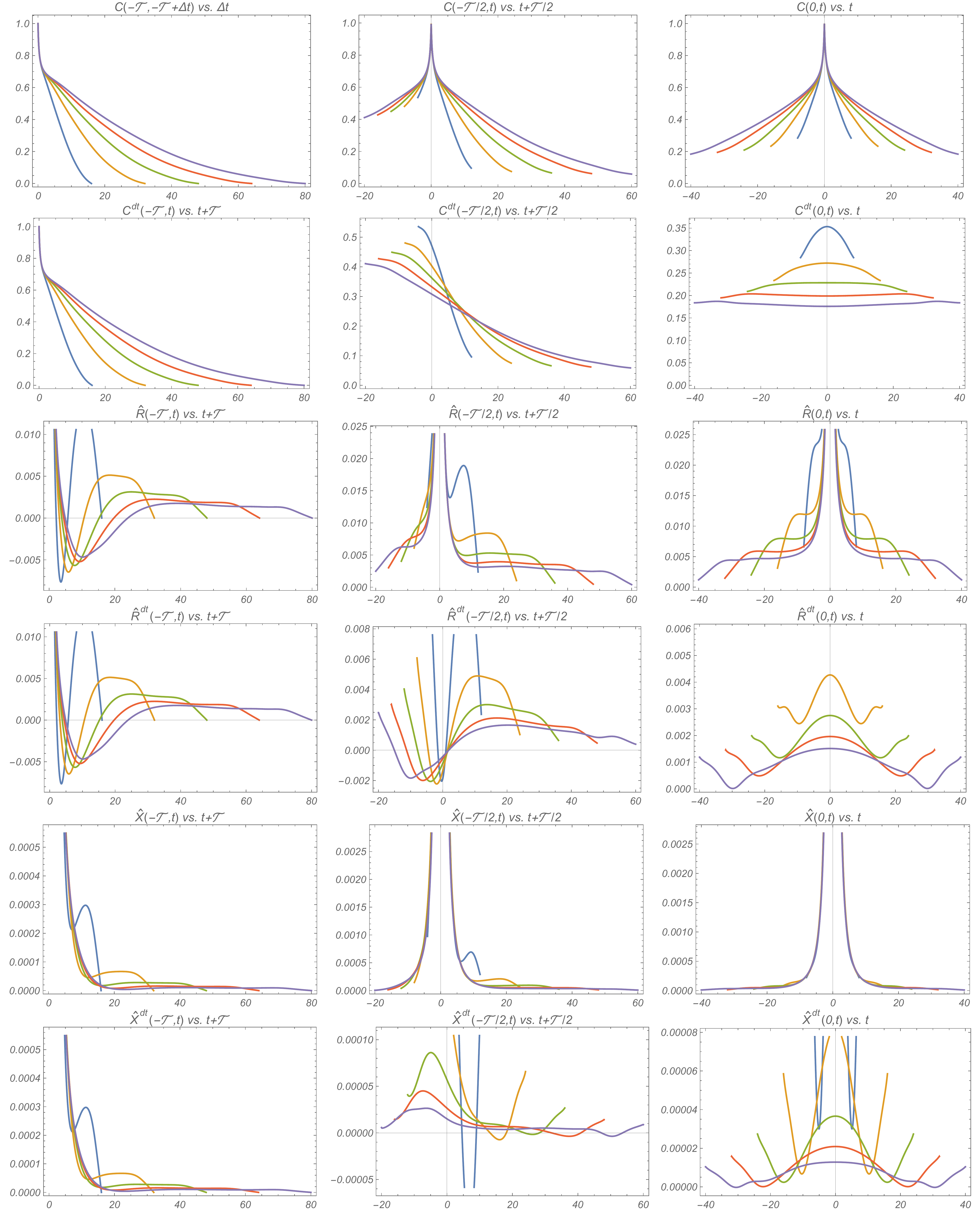} 
 \caption{Spherical $p$-Spin Glass for $\beta=1.695$ and $p=3$. Solutions for $\Tau=8 \, (blue)$, $16 \, (yellow), 24 \, (green), 32 \,(red), 40 \,(purple)$ on various strips of the $-\Tau \leq t,t' \leq \Tau$ plane. The vertical scale of the $\hR(t,t')$ and $\hX(t,t')$ functions does not allows to see the diagonal region $t -t'=O(1)$. }
\label{fig:unscaled}
\end{figure*}
In fig. (\ref{fig:unscaled}) we clearly see the features anticipated in the introduction, in particular: i) the overlap between the initial (final) configuration and the $t=0$ intermediate configuration $C(\pm \Tau,0)$ tends to a finite value in the $\Tau \rightarrow \infty$ limit, ii) the functions $\hR(t,t')$,   $\hR^{dt}(t,t')$,  $\hX(t,t')$,   $\hX^{dt}(t,t')$ decrease with increasing $\Tau$ for $|t-t'|=O(\Tau)$.

\begin{figure*}
\centering
\includegraphics[scale=.3]{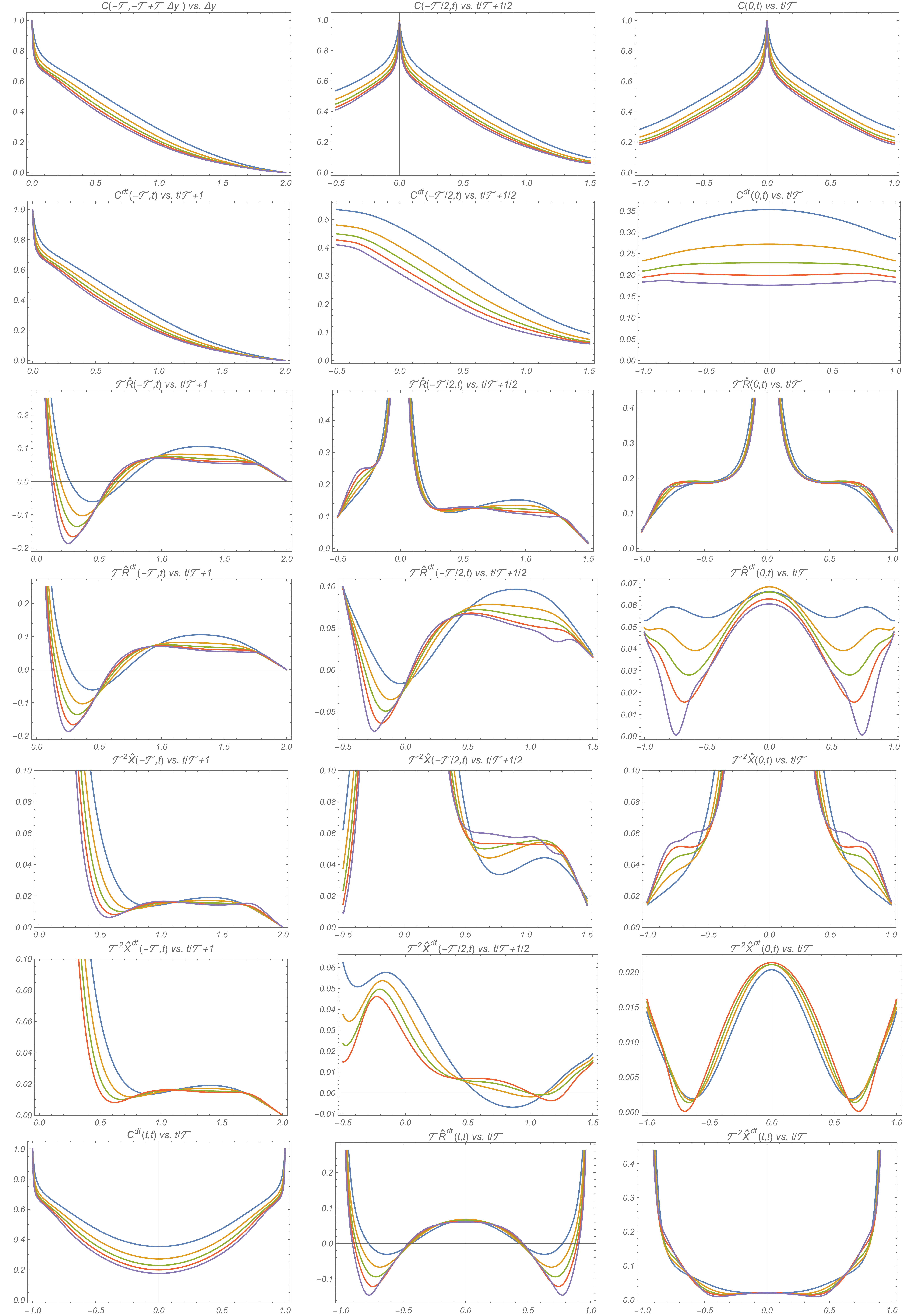} 
 \caption{Spherical $p$-Spin Glass for $\beta=1.695$ and $p=3$. Scaled solutions for $\Tau=8 \, (blue)$, $16 \, (yellow), 24 \, (green), 32 \,(red), 40 \,(purple)$ on various strips of the $-\Tau \leq t,t' \leq \Tau$ plane.}
\label{fig:scaled}
\end{figure*}
\begin{figure*}
\centering
\includegraphics[scale=.35]{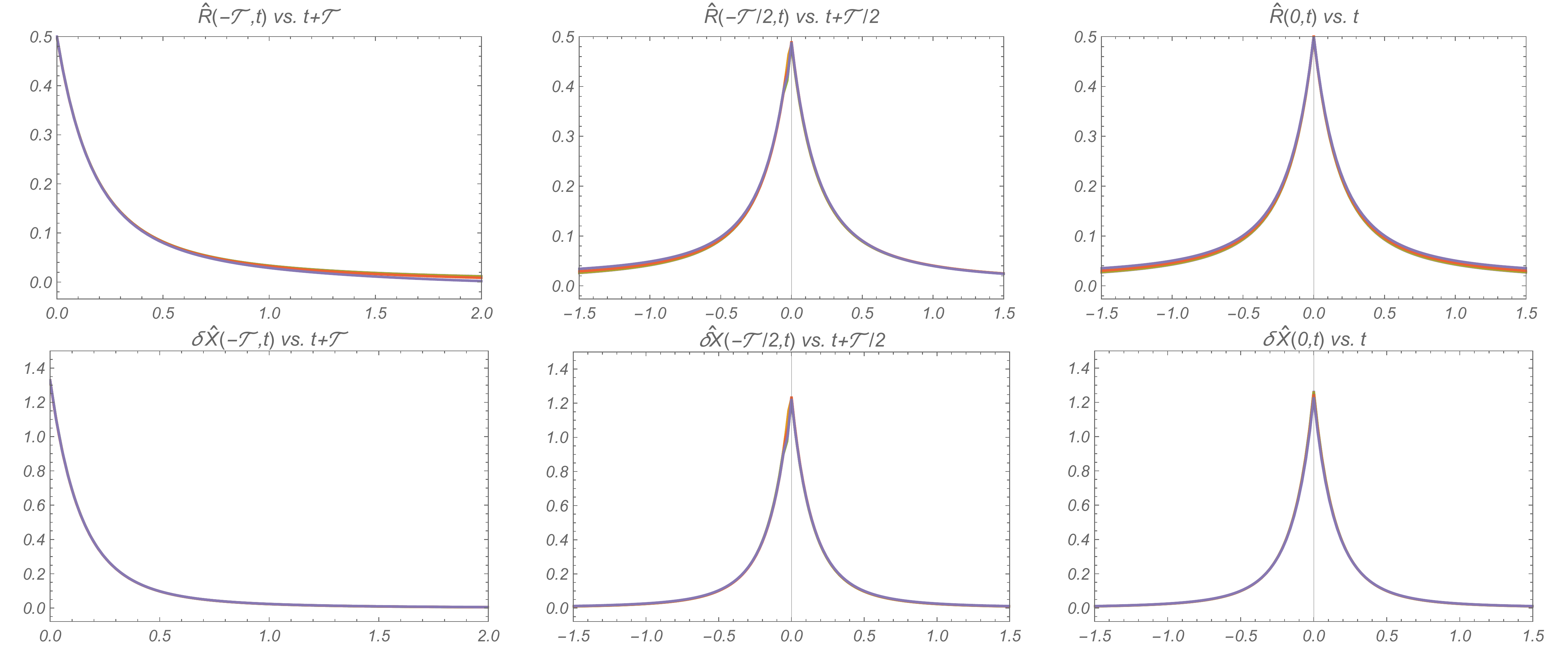} 
 \caption{Spherical $p$-Spin Glass for $\beta=1.695$ and $p=3$. The functions $\hR(t,t')$ and $\delta \hX(t,t')=\hX(t,t')+\delta(t-t')/2$  for $\Tau=8,16,24,32,40$ on various slices of the $-\Tau \leq t,t' \leq \Tau$ plane. The vertical and horizontal ranges are appropriate to visualize the diagonal region $t -t'=O(1)$. On this scale the functions for different $\Tau$'s are almost indistinguishable.}
\label{fig:unscaled-diag}
\end{figure*}
In fig.  (\ref{fig:scaled}) the same data are rescaled to fully demonstrate the asymptotic limit $\Tau \rightarrow \infty$ discussed in sec. \ref{asymptotic}. The same-trajectory functions  $\hR(t,t')$ and $\hX(t,t')$ deviate from this scaling in the region $|t-t'|=O(1)$ that goes to zero on the scale of the plots for $\Tau \rightarrow \infty$. In the region $|t-t'|=O(1)$ $\hR(t,t')$ and $\hX(t,t')$ converge to a finite limit asymptotically as shown in fig. (\ref{fig:unscaled-diag}).
One can also check that in that region $C(t,t')$, $\hR(t,t')$ and $\hX(t,t')$ satisfy the equilibrium relationships (\ref{eqR}) and (\ref{eqX}) consistently with the fact that the system is in equilibrium on scales much smaller than $\Tau$ in a field slowly varying on a $O(\Tau)$ timescale.
\begin{figure}
\centering
\includegraphics[scale=.28]{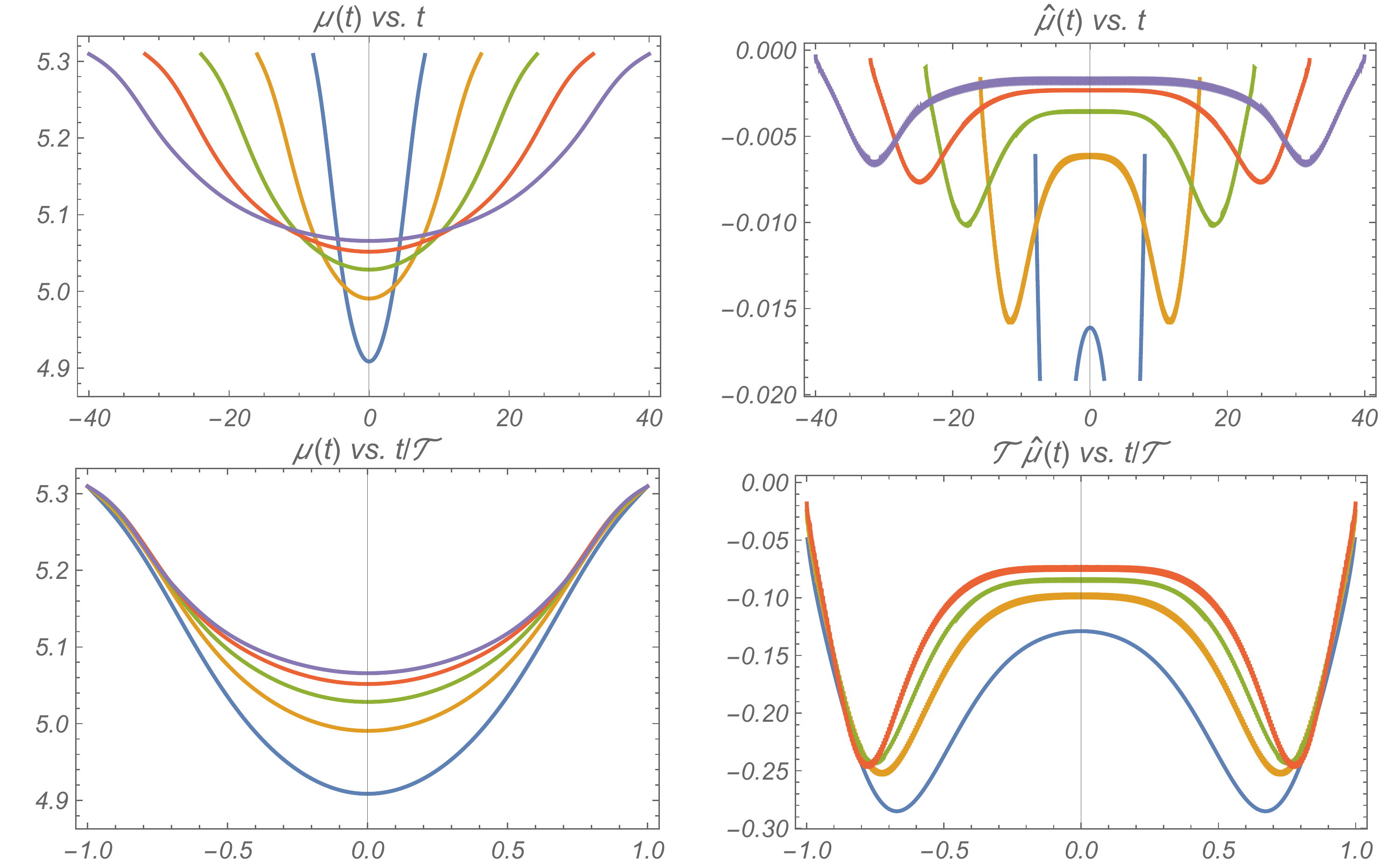} 
 \caption{Spherical $p$-Spin Glass for $\beta=1.695$ and $p=3$. The auxiliary functions $\mu(t)$ and $\hat{\mu}(t)$ for $\Tau=8 \, (blue), 16 \, (yellow), 24 \, (green), 32 \,(red), 40 \,(purple)$. The $\Tau=40$ value in the rescaled plot for $\hat{\mu}(t)$ has not been plot because the extrapolation from $N_t=2000$ seems not accurate enough on that scale.}
\label{fig:mu-mus}
\end{figure}
The auxiliary functions $\mu(t)$ and $\hat{\mu}(t)$ are shown in fig. (\ref{fig:mu-mus}). Note that  the $\Tau^{-1}$ scaling of $\hat{\mu}(t)$   is consistent with the fact that   $C(t,t')$, $\hR(t,t')$ and $\hX(t,t')$ satisfy equilibrium relationship on the diagonal according to eq. (\ref{mueq}). However $\mu(t)$ changes with time implying equilibrium on finite time-scales but not on the global scale $O(\Tau)$.
\comment{
The asymptotic scaling discussed in the main text suggests that the leading corrections are $O(1/\Tau)$. In fig. (\ref{fig:mu-extra}) we plot the data for $\mu(t)$ together with a $1/\Tau$ fit whose quality is excellent for $\Tau \geq 24$.}
\begin{figure}
\centering
\includegraphics[scale=.4]{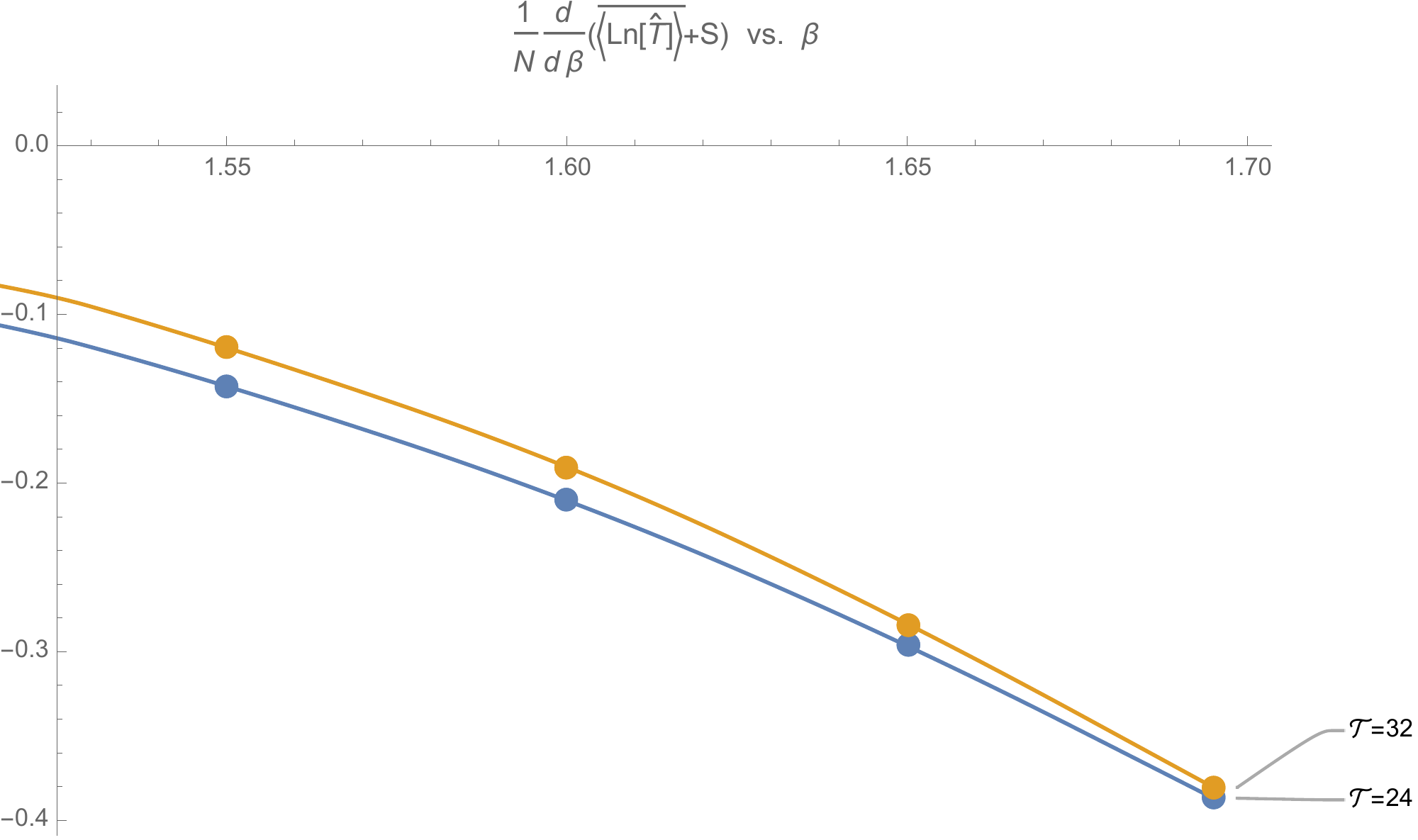} 
 \caption{Spherical $p$-Spin Glass with $p=3$. Derivative of the total transition probability with respect to the inverse temperature for $\Tau=24,32$ and $\beta=1.55$, $1.6$, $1.65$, $1.695$, the lines are second-order interpolations.}
\label{fig:der-ene}
\end{figure}
In fig. (\ref{fig:der-ene}) we plot the derivative of $\overline{\langle \ln \hT \rangle}+S$ with respect to the inverse temperature  computed according to expression (\ref{devrate}). According to the result of the previous section this quantity should go to zero in the $\Tau \rightarrow \infty$ limit for $\beta < \beta_d=1.6329$. On the other hand the asymptotic behavior sets in when $\Tau$ is larger than the equilibrium relaxation time that diverges as $|T-T_d|^{-\gamma}$  where $\gamma=1.765$ is a MCT exponent \cite{crisanti1993sphericalp,caltagirone2012critical,ferrari2012two}. Thus at any $\Tau$, no matter how large, there is always a range of temperatures $\Delta T \propto \Tau^{-1/\gamma}$  such that for $T-T_d<\Delta T$, $\overline{\langle \ln \hT \rangle}+S$ is smaller than zero. The effect decreases with increasing $\Tau$ as the figure shows but is still significant at the values of $\Tau$ that we could study. On the other hand the curves seems to have converged to $-.47$ at $\beta=1.695$ (as data from $\Tau=40$ also suggest) and this result, supplemented with the information that it must be zero asymptotically at $\beta_d$, allows for a rough estimate of the integral leading to 
\beq
\mathcal{A}=-0.014 \ \, \mathrm{for}\ \beta=1.695
\eeq

\subsection{Numerical Solution}
\label{sec:num}

\subsubsection{Newton-Krylov Methods}
To solve the equations numerically  time was discretized in steps $\Delta t= \Tau/N_t $ for integer $N_t$ up to $N_t=2000$. The six functions can then be jointly represented as  two real matrices $C$ and $C^{dt}$ of size $2(2 N_t+1)$ so that the total number of variables is $O(32 N_t^2)$.
Using standard formulas for discrete integrals and second-order derivatives (see below) one obtains expressions with an $O(\Delta t^2)$ error.

It turned out that the equations can be solved by Newton's method. To initialize the algorithm an approximate solution not too far away from the correct solution can be obtained from the analytic solutions in the free case.
One can start from $\beta=0$ at finite $\Tau$ and switch on the temperature, then at fixed temperature, $\Tau$ can be changed changing the discretization parameter $\Delta t$ at fixed $N_t$ by small amounts. At fixed $\Tau$ and $\beta$, $\Delta t$  can be reduced extrapolating the result of a coarser grid to a finer grid (larger $N_t$) and using it as a starting point for Newton's method at the new $N_t$.
The main technical problem is that every iteration of Newton's method requires to invert the Jacobian of the equations,  a $(32 N_t^2)\times (32 N_t^2)$  matrix.  Even exploiting the symmetries of the problem, with current technology {\it exact} inversion of the Jacobian becomes unfeasible for $N_t$ of the order $70-80$  limiting the values of  $\Tau$ that can be studied.
 $N_t$ could instead be increased up to $2000$ using an approximate method for the well-studied  problem \cite{saad2003iterative} of solving a very large linear system $A \, x=b$. Specifically I used the Generalized Minimal Residue (GMRES) algorithm \cite{saad1986gmres,saad2003iterative} that requires the computation of the Krylov subspace of order $k$, defined by the vectors  $\{ b, A\,b , A^2\,b,A^3\,b,\dots,A^kb\}$.
The advantage of Krylov methods is that one always work with vectors $v$ and has to perform matrix multiplications $A \, v$ without the need to store the full  $(32 N_t^2)\times (32 N_t^2)$ matrix $A$. 
In GMRES one searches for an approximate solution in the  Krylov sub-space of order $k$: an orthonormal basis is obtained using a numerically stable Gram-Schmidt orthogonalisation called Arnoldi iteration and the solution is then found by least squares minimization of the linear equations in this space.
The advantage of the method is that the error decreases systematically increasing $k$, the drawback is that it requires to generate and store all the $k$ vectors of the basis and $k$ cannot be too small to obtain accurate solutions. 
Two important ingredients that are key to the efficiency of the procedure are
preconditioning and compression to be discussed later. 
I wrote a code using Mathematica (retrievable in the ancillary files section of \cite{rizzo2020path}) being able to reach values of  $N_t=2000$  with $k=400$ using up to 90$\%$ memory on a cluster with 256 Giga of RAM. 

\begin{figure*}
\centering
\includegraphics[scale=.6]{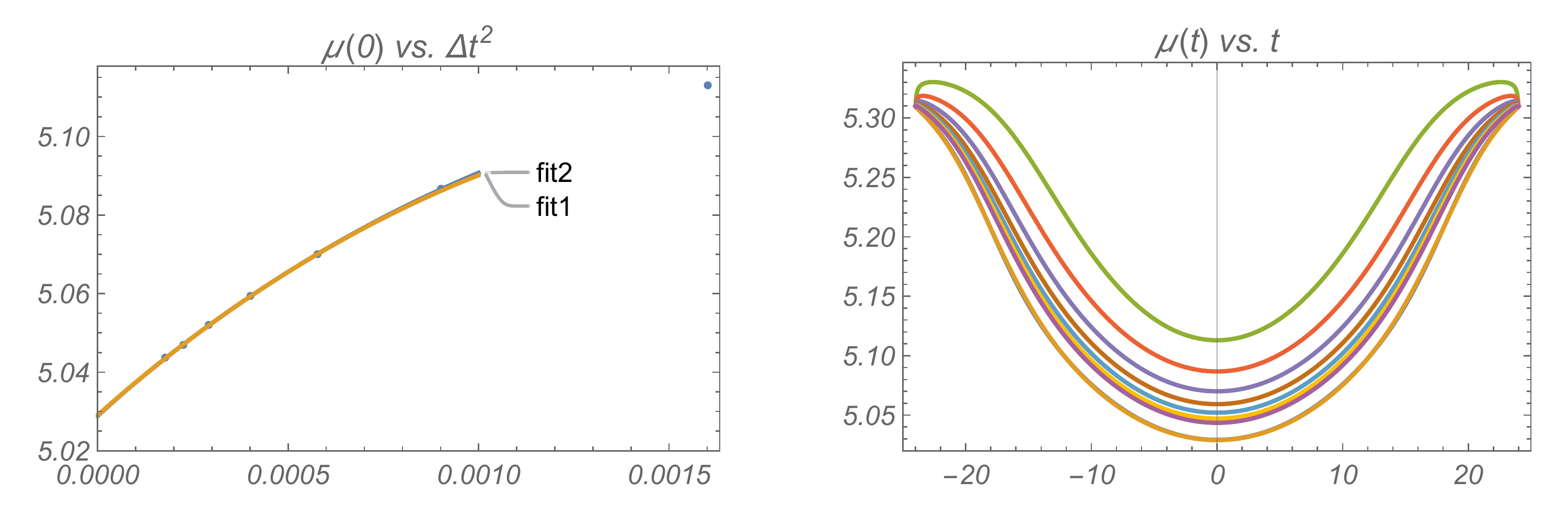} 
 \caption{Spherical $p$-Spin Glass with $p=3$, numerical solution for $\beta=1.695$, $\Tau=24$ and 
 $N_t=600$, $800$, $1000$, $1200$, $1400$, $1600$, $1800$.  Left: numerical values of $\mu(0)$, the lines are two polynomial interpolation of the form $c_0+c_2\,\Delta t^2+c_3\,\Delta t^3+c_4\,\Delta t^4$ over the data for $\{1200$,$1400$,$1600$,$1800\}$ and  $\{1000$,$1200$,$1400$,$1600\}$. The polynomial extrapolations give respectively $c_0=5.0290$ and $c_0=5.0288$. Right: from top to bottom data for increasing values of $N_t$ (individual points are not distinguishable at the scale of the plot). At the bottom there are two (indistinguishable) lines obtained from polynomial interpolations performed separately for each time $t$, one of the form $c_0+c_2\,\Delta t^2+c_3\,\Delta t^3+c_4\,\Delta t^4$ over the data for $\{1200$,$1400$,$1600$,$1800\}$, and the other of the form $c_0+c_2\,\Delta t^2+c_3\,\Delta t^3$ over the data for $\{600$,$800$,$1000\}$.}
\label{fig:Rich-mu}
\end{figure*}

\begin{figure*}
\centering
\includegraphics[scale=.6]{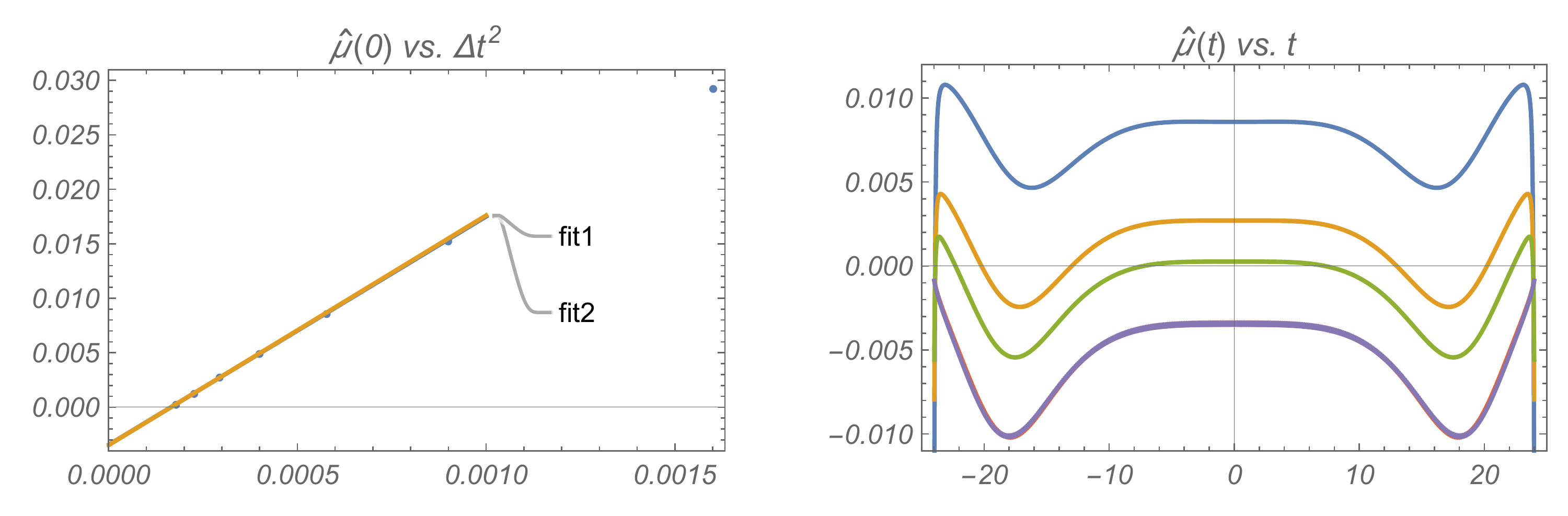} 
 \caption{Spherical $p$-Spin Glass with $p=3$, numerical solution for $\beta=1.695$, $\Tau=24$ and 
 various $N_t$.  Left: numerical values of $\hat{\mu}(0)$ for $N_t=600$, $800$, $1000$, $1200$, $1400$, $1600$, $1800$, the lines are two polynomial interpolation of the form $c_0+c_2\,\Delta t^2$ over the data for $\{1600$,$1800\}$ and  $\{1400$,$1600\}$.  Right: from top to bottom data for $N_t=1000$ (blue),  $1400$ (yellow),  $1800$ (green) (individual points are not distinguishable at the scale of the plot). At the bottom there are two (indistinguishable) lines (red, purple) obtained from polynomial interpolations of the form $c_0+c_2\,\Delta t^2$ performed separately for each time $t$,   using respectively $N_t=\{1200$,$1400\}$ and $N_t=\{1600$,$1800\}$.}
\label{fig:Rich-mut}
\end{figure*}

\subsubsection{Richardson Extrapolation}

Once the numerical solutions at fixed $\beta$ and $\Tau$ are obtained for various $N_t$, a polynomial (Richardson) extrapolation is essential to reach the $\Delta t=0$ limit and remove a few pathologies of the finite $\Delta t$ solutions. 
Let us discuss first the order of the algorithm.
For the second derivatives appearing in the equations I have used the formula 
\beq
f''(t) \approx {f(t+\Delta t)+f(t-\Delta t)-2 f(t) \over \Delta t}
\label{disc2dev}
\eeq
where
\beq
\Delta t=\Tau/N_t \, .
\eeq
The above formula has a $O(\Delta t^2)$ error if $f(t)$ has continuous derivatives up to the third order. 
 The integrals have been written as
 \beq
\int_{-\Tau}^{+\Tau} f(t)dt \approx {\Delta t \over 2} (f(-\Tau)+f(+\Tau))+\sum_{i=-N_t+1}^{N_t-1} f(i\, \Delta t) \Delta t  \ .
 \eeq
The above trapezoidal rule also has a $O(\Delta t^2)$ error for a continuous function. 
As mentioned before, the full algorithm has a $O(\Delta t^2)$ error,  however this is {\it not} trivial because the functions $C(t,t')$, $\hR(t,t')$ and $\hX(t,t')$ have discontinuous odd derivatives for $t=t'$. While the discontinuity of the first derivative is canceled by the delta function the discontinuity of the third derivative leads to a $O(\Delta t)$ error in expression (\ref{disc2dev}) on the diagonal $t=t'$. 
However on the diagonal the order of the equation is not $O(1)$ but $O(1/\Delta t)$ due to the presence of the delta function in eq. (\ref{exp2}) and of $\hX(t,t')$ (that can be written as a delta function on the diagonal plus a regular part) in eq. (\ref{exp3}), thus even if the absolute error on the diagonal is $O(\Delta t)$, the {\it relative} error is $O(\Delta t^2)$. 
In Fig. (\ref{fig:Rich-mu}) it is demonstrated that the error is indeed $O(\Delta t^2)$ for $\mu(t)$ and it is shown how that the $\Delta t \rightarrow 0$ limit is safely reached by polynomial extrapolations.
The figures discussed in the previous sections were all obtained by means of polynomial extrapolations on the largest $N_t$ set available (up to $N_t=2000$ for $\Tau=32,40$) using in most cases a fourth order form $c_0+c_2\,\Delta t^2+c_3\,\Delta t^3+c_4\,\Delta t^4$ where the vanishing of the linear term $\Delta t$ was imposed.
One should note that the finite $N_t$ curves often display pathologies due to the discretisation that tend to be less severe increasing $N_t$. For instance in the right of fig. (\ref{fig:Rich-mu}) we see that $\mu(t)$ at finite $N_t$ displays cusps close to $t=\pm \Tau$ that are absent for $N_t \rightarrow \infty$. It is impressing how a simple polynomial extrapolation over $N_t$ leads to the disappearance of these spurious features and allows to obtain accurate predictions
 with relatively small values of $N_t$ that are indistinguishable from extrapolations obtained from considerably larger values of $N_t$.
The same features are also seen in fig. (\ref{fig:Rich-mut}), in this case corrections higher that  $O(\Delta t^2)$ are so small that too high-order interpolation functions overfit that data and it is convenient to use the form $c_0+c_2\,\Delta t^2$. Besides we see that, even if the finite $N_t$ results for $\hat{\mu}(t)$ have pronounced spurious cusps close to $t \pm \Tau$ and may even have the wrong sign, the extrapolations are cuspless and negative for all $t$.

\subsubsection{Preconditioning}

Efficient Krylov methods often require preconditioning which amount to reduce the span of the eivengalues of $A$. The equations in the form (\ref{spequa}) and (\ref{spequadt}) are ill-conditioned because the operator $M$  contains second-order derivatives  leading to an unbounded continuous spectrum.
To overcome this problem I have multiplied equations  (\ref{spequa}) and (\ref{spequadt}) times $M^{-1}$. The corresponding Jacobian turns out  to have a discrete and bounded spectrum as can be seen numerically using the fact that the Arnoldi diagonalization allows to obtain an approximate set of eigenvalues and eigenvectors.
More details on the procedure can be found in the commented Mathematica codes provided in the ancillary files section of \cite{rizzo2020path}. 

\subsubsection{Compression}

To compute the equations and the matrix-times-vector products $J \, c$ involving the Jacobian $J$ we have to store huge matrices and perform multiplications and element-wise operations, for which many highly-optimized and parallelized libraries exist.
One can solve eq. (\ref{spequa}) and (\ref{spequadt}) for generic $C$ and $C^{dt}$ and verify that as expected the algorithm converges to solutions with the  required symmetries discussed in sec. \ref{subsec:theor} . However it is useful to work from the start  in the subspace of solutions with the required symmetries obtaining  a four-fold reductions of the memory required to store $C$ and $C^{dt}$ and thus the generic element of the Krylov subspace. Besides it turns out that this allows to consider a smaller Krylov subspace to obtain the same level of accuracy. Explicit use of the symmetry however requires an efficient procedure to quickly compress and decompress the large matrices $C$ and $C^{dt}$  and the elements of the Krylov subspace.
The choice depends on the specific linear algebra programming tool used.

\section{Concluding Perspectives}
\label{sec:conclu}

In this work I have shown how the transition rate can be computed in mean-field models. As stated in the introduction, the main motivation to perform such a computation is to obtain dynamical quantitative predictions not accessible through static methods. Besides the most interesting physical outcome of the computation is to shed new light on the role of the complex landscape for ergodicity restoring processes as discussed in section \ref{sub:ergres}: 
 the conclusion is that to visit different equilibrium states it is more convenient to make intermediate jumps to off-equilibrium metastable states rather than direct jumps from one equilibrium state to another.

The methods discussed here can be applied to other mean-field systems, in the introduction we mentioned supercooled liquids models in high dimensions but the Langevin equation can be also viewed as an algorithm in the wider family of (stochastic) gradient-descent algorithms that are widely used in the context of machine-learning and statistical inference.
The analysis of the algorithm performance in the context of high-dimensional inference has been initiated in \cite{mannelli2020marvels}
unveiling a glassy phase that limits the efficiency of the algorithm. Those results where obtained in the thermodynamic limit where the evolution of the algorithm can be associated to relaxational dynamic equations similar to those of the spherical $p$-spin SG model \cite{crisanti1993sphericalp,cugliandolo1993analytical} recently reconsidered in \cite{folena2020rethinking}. On the hand other realistic problems have {\it finite} size and thus the algorithm can overcome those thresholds albeit with larger convergence times that can be studied by the methods presented here switching from the initial value dynamical equations studied in \cite{mannelli2020marvels} to boundary value dynamical equations.
Concerning the extension to supercooled liquid in high dimensions, the equations can be derived and solved in principle by the GMRES algorithm although the actual implementation is likely to be considerably more complex than for the spherical model \cite{manacorda2020numerical}.
In this respect it is worth noticing that for more complex problems one can use a Jacobian-free method, in which  the product $J \, v$ required by Newton-Krylov methods is approximated by $E(c+\epsilon\, v)-E(c))/\epsilon$ for small $\epsilon$ and thus  only the numerical computation of the equations $E(c)$ is required.
Furthermore other algorithms exist that do not require to store the whole  Krylov Space, {\it e.g.} Biconjugated Gradient \cite{saad2003iterative}; in general they are less safe and controlled than GMRES, but an efficient algorithm requiring less memory would be welcomed. Note that GMRES  allows to have an approximation for the spectrum of the true Jacobian,  which could be useful information to devise alternative algorithms.

We have mentioned that in the {\it asymptotic} regime $\Tau \rightarrow \infty$ the  solutions solve $\Tau$-independent scaling equations with  the time derivatives dropped.
In the context of off-equilibrium dynamics the same thing happens and allows to establish a remarkable connection with the equations of the static/replica framework  \cite{cugliandolo1993analytical}. This grants that the phase diagram and many non-trivial off-equilibrium quantities can be determined without solving explicitly the dynamical equations and it would be interesting to determine if there is also some static potential from which the results of the $\Tau \rightarrow \infty$ limit can be recovered without solving the dynamics. A natural question is weather such a static potential is the celebrated Franz-Parisi potential \cite{franz1997phase} that develops a secondary minimum at $T_d$. One should indeed remember that, while the minimum is associated to equilibrium states and configurational complexity, it is not known if the potential {\it difference} between the secondary minimum and the maximum  is in fact  associated  to any quantity computed by actual dynamical methods as those considered here.
 
An interesting open question is how one should expect the solutions to look like at $T_d$, in particular in the asymptotic form. Some guidance could be offered by the fact that in the $T \rightarrow T_d$ limit the results should match somehow those obtained in \cite{rizzo2014long,rizzo2015qualitative,rizzo2016dynamical,rizzo2016glass,rizzo2020solvable} by considering finite-size/finite-dimensional systems directly at $T=T_d$ (from a technical point of view the two approaches differ by which one of the limits $T \rightarrow T_d$ or $N \rightarrow \infty$ is taken first). This suggests for instance that the important physics occurs close to the plateau value of the overlap.

It should be noted that the dynamical equations have been solved by making a Replica-Symmetric ansatz on the $n$ dynamical replicas introduced to compute the logarithm of the rate (see eq. (\ref{RSansatz}) in appendix \ref{sec:action}) and one may ask if it yields correct results or a RSB ansatz should be used. While a full answer requires a complicated analysis of the stability of the saddle-point equations,  an instability could nonetheless manifests itself in some inconsistent value of some physical quantity, much as the RS entropy of the Sherrington-Kirkpatrick model becomes negative at low temperatures \cite{mezard1987spin}. This is not the case as the numerical solution did not show any visible inconsistency  suggesting that the RS ansatz is correct in the range of temperatures considered.

To conclude we recall that in section \ref{sub:ergres} it was shown that the presence of many metastable states implies that  the {\it exponentially small} probability that the system jumps to another equilibrium state in a {\it finite} time is not trivially related to 
 the {\it exponentially large} time-scale $\tau_{erg}$ over which the system jumps to another equilibrium state with {\it finite} probability. A question worth of further investigation is weather the latter could also be computed by similar methods.

\begin{acknowledgments}
I acknowledge the financial support of the Simons Foundation (Grant No. 454949, Giorgio Parisi).
I thank E. Zaccarelli for substantial help with computing resources.
 \end{acknowledgments}

 \appendix

\section{Path Integral Expression of the Rate}

\subsection{Replicas}

The logarithm in the quenched average can be eliminated by the Replica method
\beq
\sum_{\sigma,\tau} P_{eq}(\sigma)P_{eq}(\tau)\, \ln \hT(\sigma,\tau) \ ,
\eeq
 furthermore we consider systems with quenched disorder whose equilibrium properties can also be studied by the replica method, thus we introduce the following object:
\begin{widetext}
\beq
Z_{\hT}(m,m',n) \equiv \sum_{\sigma_1 \dots \sigma_m}\sum_{\tau_1 \dots \tau_{m'}}\exp\left[-\beta \sum_{i=1}^m H_J(\sigma_i)-\beta \sum_{i=1}^{m'} H_J(\tau_i)\right]\hT(\sigma_1,\tau_1)^n
\eeq
and we have:
\beq
\sum_{\sigma,\tau} P_{eq}(\sigma)P_{eq}(\tau)\, \ln \hT(\sigma,\tau)= \lim_{m \rightarrow 0} \lim_{m' \rightarrow 0}\lim_{n \rightarrow 0}{d \over dn}  \ln Z_{\hT}(m,m',n)\ .
\eeq 
As usual in the context of mean-field models the thermodynamic limit is taken before the above limits. Note that, the quantity $\lim_{m \rightarrow 0, m' \rightarrow 0} \,Z_{\hT}(m,m',n)$ at finite $n$ allows to study the large deviations of $\ln \hT(\sigma,\tau)$ and determine if it is self-averaging with respect to the equilibrium configurations $\sigma$ and $\tau$.
As we will see in the next subsection it is possible to obtain an integral representation of the dynamics (see \cite{Zinn-Justin2002}, chapter 17):
\beq
\hT(\sigma_1,\tau_1)=\int_{s(-\Tau)=\sigma_1,s(\Tau)=\tau_1} d s\exp\left[{1 \over 2} \int d1 d2 s(1)\Gamma(1,2)s(2)- \beta \int d1 H_J(s(1)) \right]\ .
\eeq
Where the $1$ and $2$ are coordinates that collect a time variable and Fermionic variables introduced to obtain the integral representation.
The expression is made of two parts: a universal dynamic one encoded by the matrix $\Gamma(1,2)$ and an interaction part that depends on the model encoded in the Hamiltonian $H_J$.
The above expressions leads to:
\beqa
Z_{\hT}(m,m',n)  & = & \sum_{\sigma_1 \dots \sigma_m}\sum_{\tau_1 \dots \tau_{m'}}
\prod_{a=1}^n \left[ \int_{s_a(-\Tau)=\sigma_1,s_a(\Tau)=\tau_1}   [ds_a]\right] \exp\left[ \sum_{a=1}^n {1 \over 2} \int d1 d2 s_a(1)\Gamma(1,2)s_a(2)+ \right. 
\nonumber
\\
& & \left.-\beta \sum_{i=1}^m H_J(\sigma_i)-\beta \sum_{i=1}^{m'} H_J(\tau_i) -\beta \sum_{a=1}^n \int d1 H_J(s_a(1)) \right] \ ,
\eeqa
this in turn can be written in a compact form as:
\beq
Z_{\hT}(m,m',n)  =\int  [d { \bf s}]\exp\left[ {1 \over 2}\int {\bf d1 d2 s(1)\Gamma(1,2)s(2)}- \beta \int d{\bf 1} H_J({\bf s(1)}) \right]
\label{genz}
\eeq
\end{widetext}
where the  bold index ${\bf 1}$ runs over the replicas and the dynamical indexes
\beq
{\bf s(1)}=\{ \sigma_1, \dots, \sigma_m, s_1(1), \dots, s_n(1),\tau_1, \dots, \tau_{m'} \}
\label{gens}
\eeq
and we have:
\beq
\int f({\bf s(1)})d{\bf 1} = \sum_{i=1}^m f(\sigma_i)+ \sum_{i=1}^{m'} f(\tau_i) + \sum_{a=1}^n \int d1 f(s_a(1)) \ .
\eeq
The dynamical operator ${\bf \Gamma(1,2)}$ is diagonal with respect to the replica indexes and associates each dynamical replica with the boundary conditions at $\sigma_1$ and $\tau_1$.  Its definition in the specific representation we will use is given in eq. \ref{Gamma12}.
As usual the great advantage of having the above compact representation is that one can perform the average of the disorder and then perform standard manipulations yielding an expression formally identical to the one obtained in the case of a static replica computation.

\subsection{Path Integral Representation of Langevin Dynamics}
\label{sec:pathint}

This section discusses the path integral representation of Langevin dynamics, given in terms of a single component $q$.  The one-dimensional case is discussed for simplicity, the generalisation to $N$-dimensional vectors used in the paper is straightforward.
The Langevin equation reads
\beq
{1 \over \Gamma_0}\dot{q}=-\beta {d H \over dq}+\xi\, ,\  \langle \xi(t)\xi(t')\rangle={2 \over \Gamma_0}\delta(t-t') 
\eeq
and we discretize it as:
\beq
{1 \over \Gamma_0}{q_{i+i}-q_i \over \Delta t}=-\beta \left( c\,{d H \over dq_i}+(1-c){d H \over dq_{i+1}} \right)+\xi_i
\eeq
where $0 \leq c \leq 1$ is an arbitrary constant: in It\^o discretization with have $c=1$, in Stratonovich we have $c=1/2$.
Enforcing the equations through an integral representation we can write the average over trajectories at fixed initial and final conditions  as an integral
\beqd
T(\sigma|\tau)=\int \left( \prod_{i=1}^{N-1}dq_i \right) \left( \prod_{i=0}^{N-1} {d\hat{q}_i \over 2 \pi} \right)
\times
\eeqd
\beqd
\times \exp\left[-\sum_{i=0}^{N-1}\left( \hat{q}_i E_i \Delta t -\ln\left[ {1 \over \Gamma_0}-(1-c)\beta {d^2 H \over dq_{i+1}^2}  \, \Delta t \right]\right)\right]
\eeqd
where the interval between the initial and final time is divided in $N$ sub-intervals of size $\Delta t$ and:
\beq
E_i \equiv {1 \over \Gamma_0}{q_{i+i}-q_i \over \Delta t}+\beta \left( c\,{d H \over dq_i}+(1-c){d H \over dq_{i+1}} \right)-\xi_i
\eeq
The logarithm comes from the determinant of the Jacobian that gets contribution only from the diagonal since the Jacobian is a triangular matrix.
Expanding the Jacobian at first order in $\Delta t$ we obtain:
\beq
T(\sigma|\tau)=\int \left( \prod_{i=1}^{N-1}dq_i \right) \left( \prod_{i=0}^{N-1} {d\hat{q}_i \over 2 \pi \Gamma_0} \right) \exp[-\mathcal{L}]
\eeq
The Lagrangian reads:
\beq
\mathcal{L}=\sum_{i=0}^{N-1}\left( \hat{q}_i E_i \Delta t -(1-c)\beta {d^2 H \over dq_{i+1}^2} \, \Gamma_0 \, \Delta t \right) 
\eeq
and in the continuum limit we have
\beq
{\mathcal L}=\int dt \left[{1 \over \Gamma_0}\left(\hat{q}\dot{q}-\hat{q}^2\right) +\hat{q} \beta {d H \over dq}- (1-c) \Gamma_0 \beta {d^2H \over dq^2} \right]\ .
\eeq
Note that unexpectedly the continuum limit expression depends on the microscopic parameter $c$ of the discretization while one would expect it to be irrelevant. This is a well-known ambiguity of path integral representation of stochastic equations. One can choose to use the It\^o discretization corresponding to $c=1$ and neglect it but it will resurface later in the computation.
In the following we prefer to keep it also to remind us that the continuum limit of stochastic equations must be taken with care as the ordinary rules of calculus (integration by parts, differentiation, chain rules) are modified.
Beside we will use Hamiltonians where the interaction part is just linear (the $p$-spin interactions) and thus in the end we will go back to special Langevin equation for a single variable.
Let us consider the symmetric rate defined as
\beq
\hT(\sigma,\tau) \equiv T(\sigma|\tau)e^{{\beta \over 2}\left( H(\sigma)-H(\tau)\right)}
\eeq
In the continuum limit we would expect the following to be an equality
\beq
{\beta \over 2}\left( H(\sigma)-H(\tau)\right) \neq {\beta \over 2}\int_{-\Tau}^{+\Tau}dt \, \dot{q} {\partial H \over dq} \ .
\eeq
Instead  in order to get the correct expression  we should go back to the discretized expression. We have:
\beq
 H(q_{i+1})-H(q_i)= {d H \over dq_i} \Delta q_i+  {1 \over 2}{d^2H \over dq_i^2}\Delta q_i^2
\eeq
then we have to use the fact that in the Lagrangian we use the following discretized definition of $dH/dq$ 
\beq
 c\,{d H \over dq_i}+(1-c){d H \over dq_{i+1}} \ .
\eeq
By rewriting the differential as
\beq
{d H \over dq_i}=\left( c\,{d H \over dq_i}+(1-c){d H \over dq_{i+1}} \right)-(1-c){d^2 H \over dq_{i}^2}\Delta q_i
\eeq
we obtain
\beq
d H = {d H \over d q} dq+\left(c-{1 \over 2}\right){d^2 H \over dq^2} dq^2 \ .
\eeq
We can see that for $c=1$ we recover It\^o's lemma while for the Stratonovich prescription $c=1/2$ we find that the ordinary chain rule applies.
The second term cannot be neglected because it gives an $O(dt)$ contribution but we can make the replacement
\beq
d q^2= 2 \, \Gamma_0 \,dt
\eeq 
and obtain
\beqd
{\beta \over 2}\left( H(\sigma)-H(\tau)\right)={\beta \over 2}\int_{-\Tau}^{+\Tau} dt \left(  {d H \over d q} \dot{q}+\left(c-{1 \over 2}\right){d^2 H \over dq^2} 2 \, \Gamma_0 \right)\ .
\eeqd
The same result can also be obtained reabsorbing the term $\Delta q^2$ in the term $\dot{q}^2$ below (see {\it e.g.}  \cite{Zinn-Justin2002}, section 4.6).
Making the following change of variable 
\beq
\hat{q}=\hat{x}+{\dot{q} \over 2}
\eeq
we finally obtain
\beq
\hat{\mathcal{L}}(t)={1 \over \Gamma_0}\left({\dot{q}^2 \over 4}-\hat{x}^2 \right)+\hat{x}\beta {d H \over dq}-{1 \over 2}\Gamma_0 \beta {d^2H \over dq^2}\ .
\eeq
We can further integrate out the $\hat{x}$:
\beq
\hat{\mathcal{L}}'(t)={1 \over \Gamma_0}{\dot{q}^2 \over 4}+{\Gamma_0 \over 4}\left(\beta {d H \over dq}\right)^2-{1 \over 2}\Gamma_0 \beta {d^2H \over dq^2} \ ,
\eeq
see also \cite{Zinn-Justin2002}, pag. 70.
Note that the integration over $\hat{x}$ leads to a divergent prefactor to the above path integral:
\beq
{1 \over 2 \pi \Gamma_0}\int d\hat{x}e^{\hat{x}^2 \over \Gamma_0}dt={1 \over 2 \pi \Gamma_0} \left(2 \pi {\Gamma_0 \over 2\, dt}\right)^{1/2} \ .
\eeq
which is usually buried into the  expression $[dq]$ defined as:
\beq
[dq] \equiv dq \, \left( {1/(2\Gamma_0) \over 2 \pi dt}\right)^{N/2}
\eeq
see eq. 2.20 in \cite{Zinn-Justin2002}.

\subsection{Path Integral Representation for Models with Multi-Linear Interactions}
\label{sec:pimulti}

Typical mean-field SG models have interactions that are multi-linear, while the non-linear part of the Hamiltonian is local and does not need to be decoupled in order to obtain a saddle point expression. This allows to use simplified integral representations of the dynamics in which fewer variables are introduced with respect to the general case discussed in \cite{Zinn-Justin2002}, chapter 17.
In the following we introduce a bosonic variable $\eta$ that behaves as the product of two Grassmann variables, that is we have:
\beq
\eta^2=0 \, , \ \int d\eta =0\, , \  \int \eta \, d\eta =1 
\eeq 
with it we define a new coordinate $1 \equiv (t, \eta)$ and field $s(1)$: 
\beq
s_i(1) \equiv s_i(t)+ \hat{x}_i(t) \eta \ .
\eeq
With the above definitions the interacting part of the multi-linear Hamiltonian can be written as:
\beq
\beta \int dt \sum_i \hat{x}_i(t) {d H \over d s_i}= \beta \int d1 H(s(1)) \ .
\eeq
A similar formulation is also useful  in the spherical model. In this case however the non-linear part of the Hamiltonian is due to the spherical constraint which is not local and must be treated appropriately.
Let us consider  the problem of the integral representation of Langevin dynamics of $N$ real spins $s_i$ constrained on a  $(N-1)$-dimensional surface specified by some condition $G(s)=0$. The statics of the problem can be written as
\beq
\int d^N s \,\delta(G)|\nabla G|e^{-\beta H(s)}
\eeq
A convenient way to define Langevin dynamics on the surface is to relax the delta function replacing it with a Gaussian of infinitesimal variance $\epsilon$. Then one have to compute the standard dynamical integral in presence of a Hamiltonian $H(x)=G^2(x)/(2 N \epsilon)$. For the spherical constraint on $N$ continuous spins $s_i$ we have
\beq
G(s)\equiv \sum_{i=1}^N  s_i^2-N
\eeq
the term $|\nabla G|$ is exactly equal to $N$ and can be ignored, For $H(x)=G^2(x)/(2 N \epsilon)$ we have:
\beq
\sum_i\hat{x}_i\beta {d H \over dq_i}-{1 \over 2}\Gamma_0 \beta\sum_i {d^2H \over dq_i^2}  = {\beta \over N \epsilon } G(s) \left(2  \sum_i \hat{x}_i s_i -\Gamma_0 N \right)
\eeq
plus $o(N)$ terms. The expression can be decoupled through a Hubbard-Stratonovich transformation in terms of two additional fields $\mu(t)$ and $\tilde{\mu}(t)$ and taking the limit $\epsilon \rightarrow 0$ the quadratic part disappears leading to the following  contribution to the action:
\beqd
-{1 \over 2}\int dt \hat{\mu}(t)\left( \sum_i s_i^2-N\right)-{1 \over 2}\int dt {\mu}(t)\left(2  \sum_i \hat{x}_i s_i -\Gamma_0 N \right)\ .
\eeqd
Introducing the variable $\mu(1)=\mu(t)+\hat{\mu}(t) \eta$ we can write the above term as
\beqd
-{1 \over 2}\int d1 \mu(1) \left(\sum_i s_i^2(1)- N\right)+{1 \over 2}\Gamma_0 \int dt  \mu(t)
\eeqd
Note that presence on the second term that is essential to get the correct saddle-point equations and the correct value of the rate.

\section{The $p$-spin Spherical Model: The Action and Saddle-Point Equations}
\label{sec:action}

In the standard treatment \cite{mezard1987spin} the expression of the free energy of the fully-connected spin-glass models is obtained through a number of steps.  First $n$ replicas of the model are introduced, then the disorder average over the partition function of the replicated system is performed and this leads to an expression in which different replicas are coupled. Then a Hubbard-Stratonovich transformation is performed leading to an action depending on a single local replicated spin variable and on two $n \times n$ matrices $Q_{ab}$ and $\Lambda_{ab}$. Finally the action has to be integrated over $Q_{ab}$ and $\Lambda_{ab}$ but the mean-field nature of the model allows to use the saddle-point method.
The very same manipulations can be applied to the interaction part of expression (\ref{genz}), the only difference is that instead of having spins $s_a$ labeled by indexes $a=1,\dots,n$ the spins depend on a more complex index (or coordinate) ${\bf 1}$ specified in eq. (\ref{gens}).
As a result the objects $Q$ and $\Lambda$ are (formally) matrices in this more complex coordinate.  
In spherical models the spin-variables can be also integrated out and 
 we obtain the following expression for $Z_{\hT}$ of the $p$-spin spherical model:
\begin{widetext}
\beq
\exp\, N \left[  {\beta^2 \over 4}\int d{\bf a}d{\bf b}f({\bf Q(a,b)})  + {1 \over 2} {\bf \Lambda (a, b) Q(a,b)} -{1 \over 2} \Tr \ln ({\bf \Gamma} + {\bf \mu} +{\bf \Lambda}) +const.+{1 \over 2 } \int d{\bf a} \,   \mathbf{ \mu (a)}+\sum_{a=1}^n{\Gamma_0 \over 2} \int   \mu_a(t)dt \right]
\label{logtranspher}
\eeq
\end{widetext}
The expression $const.$ above collects a number of terms coming from the explicit Gaussian integration, it is divergent and cancels the divergences associated to the expression  $\Tr \ln ({\bf \Gamma} + {\bf \mu} +{\bf \Lambda})$. These are pathologies of the path integral representation that are fixed going back to the discrete times and will be further discussed in appendix \ref{sec:trfree}. 
Note that the application of the standard manipulations to the form (\ref{genz}) would lead to the above expression without the last term that  appears instead if we want to use the simplified formulation which is suitable for multi-linear interactions.
The above expression has to be extremized with respect to ${\bf Q(ab)} $, ${\bf \Lambda(ab)}$ and with respect to $\bm{ \mu(a)}$. Extremization with respect to  ${\bf Q(ab)} $, $\bm{\Lambda(ab)}$ leads to the saddle point equations:
\beqa
{\bf \Lambda(ab)} & = & -{\beta^2 \over 2}f'({\bf Q(ab)})
\\
{\bf Q} & = & {1 \over \bm{\Gamma + \Lambda+\mu}}
\eeqa
The last equation can be rewritten as:
\beq
\bm{\Gamma(ac) Q(cb)+ \bm{\mu(a)\, \delta(ac) Q(cb)}+\Lambda(ac) Q(cb)}=\bm{\delta(ab)}
\label{fuleq}
\eeq
where there integration over the variable $\bm{c}$ is implicit.
The above expression is extremely compact, in the following we wil see that it encodes eight integro-differential equations.

From now on we specialize to the case of a 1RSB transition.
We start noticing that the initial and final configuration are weighted with the equilibrium Gibbs measure and their properties must not depend on the dynamics. This is granted by the fact that the terms depending on the dynamics in the equations for the replicas of the initial and final configurations are $O(n)$ and disappear in the $n \rightarrow 0$ limit (that must be taken first). 
This means that $Q(ab)$ with $a$ and $b$ corresponding to the $a,b =1,\dots,m+m'$ replicas associated to the equilibrium boundary conditions is the ordinary equilibrium replica matrix.
For simplicity we will work in zero field and zero random field meaning that:
\beq
f'(0)=0 \, ,
\eeq
 in this way the overlap between different equilibrium configurations is zero. Therefore above $T_d$  the solution is $Q(ab)=\delta(ab)$ that plugged into the above equations leads to:
\beq
\mu=1+{\beta^2 \over 2}f'(1)\ .
\eeq 
Below $T_d$  the solution is actually 1RSB. The replicas determining the boundary conditions should naturally belong to different RSB blocks ensuring that we are studying the transition between different states. On the other hand the equations for the dynamical part will have a non-vanishing correlation with the remaining  $x-1$ replicas in the block of the initial configuration and with the $x-1$ replicas in the block of the  final configuration.
This will lead to a correction of order $x-1$ to the dynamical equations valid at high temperature. However for $T_s<T<T_d$ we have exactly $x=1$  and thus this contribution vanishes. Therefore we will safely use the   high temperature Replica-Symmetric (RS) solution $Q(ab)=\delta(ab)$ also in the region  $T_s<T<T_d$. This is consistent with the known result that annealed and quenched averages are equivalent above $T_s$. 

The global order parameter $\bm{Q(ab)}$ can be divided into a static part corresponding to the $m+m'$ replicas, a dynamic part describing the $n$ replicas of the dynamics and a mixed part. We have seen before that, as it should, the static part does not depend on the dynamics part due to the $n \rightarrow 0$ limit.
We now focus on the dynamic part. We will make a RS ansatz on the $n$ dynamical replicas, therefore the dynamical component of $\bm{Q(ab)}$ will be characterized by two matrices 
\beq
\bm{Q(ab)} = \delta_{\alpha\beta}Q(ab)+(1-\delta_{\alpha\beta})Q^{dt}(ab)
\label{RSansatz}
\eeq
where we have moved from the full coordinates $\bm{(ab)}$ to purely dynamical coordinates $(ab)$ and replica coordinates $\alpha,\beta=1,\dots,n$. The RS ansatz implies also 
\beq
\mu_\alpha(t)= \mu(t)\, , \ \hat{\mu}_\a(t)=\hat{\mu}(t)\, 
\eeq
and we have:
\beqa
\bm{\Gamma(ab)} & = & \delta_{\alpha \beta} \Gamma(ab)
\\
\bm{\delta(ab)} & = & \delta_{\alpha \beta} \delta(ab) \ .
\eeqa
The dynamical components of equations  ({\ref{fuleq}})  can then be rewritten in a form that can be analitically continued to real values of the replica number $n$
\beq
 \Gamma(ac) Q(cb)+\beta \mu(a)\delta(ac)Q(cb)+(\Lambda Q)^{st}  =  \delta(ab)
\label{eq}
\eeq
\beq
 \Gamma(ac) Q^{dt}(cb)+\beta \mu(a)\delta(ac)Q^{dt}(cb)+(\Lambda Q)^{dt}  = 0
\label{eqdt}
\eeq
where again the integration over the variable $c$ is implicit and 
\beqa
(\Lambda Q)^{st} & = & \Lambda(ac) Q(cb)+(n-1)\Lambda^{dt}(ac) Q^{dt}(cb)+
\nonumber
\\
& + & \Lambda(a1)Q(1b)+\Lambda(a2)Q(2b) \ '
\\
(\Lambda Q)^{dt} & = & \Lambda^{dt}(ac) Q(cb)+\Lambda(ac) Q^{dt}(cb)+
\nonumber
\\
& + &(n-2)\Lambda^{dt}(ac) Q^{dt}(cb)+
\nonumber
\\
&+&\Lambda(a1)Q(1b)+\Lambda(a2)Q(2b) \ ,
\eeqa
where $1$ and $2$ label the two static replicas whose configurations are chosen as initial and final condition of the dynamics at times $\mp \Tau$. Note that the corresponding terms appear when we integrate over the full coordinate $ {\bm c}$  in eq. (\ref{fuleq}).
In the above equations we have naturally:
\beqa
 \Lambda(ab) & = & -{\beta^2 \over 2}f'(Q(ab))
\\
\Lambda^{dt}(ab) & = & -{\beta^2 \over 2}f'(Q^{dt}(ab))\ .
\eeqa
$Q(ab)$ and $Q^{dt}(ab)$ can be expressed in terms of four two-time functions as:
\beqa
Q(ab) & \equiv &  C(t_a,t_b)+\hat{R}_{1}(t_a,t_b)\eta_a+\hat{R}_{2}(t_a,t_b)\eta_b+
\nonumber
\\
&+&\hat{X}(t_a,t_b)\eta_a \eta_b 
\\
Q^{dt}(ab) &  \equiv & C^{dt}(t_a,t_b)+\hat{R}^{dt}_{1}(t_a,t_b)\eta_a+\hat{R}^{dt}_{2}(t_a,t_b)\eta_b+
\nonumber
\\
&+&\hat{X}^{dt}(t_a,t_b)\eta_a \eta_b 
\eeqa
The same representation can be obtained from any function $A(ab)$ as
\beqa
A(ab) & \equiv &  C_A(t_a,t_b)+\hat{R}_{1,A}(t_a,t_b)\eta_a+\hat{R}_{2,A}(t_a,t_b)\eta_b+
\nonumber
\\
&+&\hat{X}_A(t_a,t_b)\eta_a \eta_b \ .
\eeqa
from which we obtain
\beqa
C_{\Lambda}(t_a,t_b) & = &-{\beta^2 \over 2}f'(C(t_a,t_b))
\label{def1}
\\
\hat{R}_{1,\Lambda}(t_a,t_b) & = &-{\beta^2 \over 2}f''(C(t_a,t_b))\, \hat{R}_{1}(t_a,t_b)
\label{def2}
\\
\hat{R}_{2,\Lambda}(t_a,t_b) & = &-{\beta^2 \over 2}f''(C(t_a,t_b))\, \hat{R}_{2}(t_a,t_b)
\label{def3}
\eeqa
\beqd
\hat{X}_{\Lambda}(t_a,t_b)  = -{\beta^2 \over 2}f''(C(t_a,t_b))\, \hat{X}(t_a,t_b)+
\eeqd
\beq
-{\beta^2 \over 2}f'''(C(t_a,t_b))\, \hat{R}_{1}(t_a,t_b)\hat{R}_{2}(t_a,t_b)
\label{def4}
\eeq
\beq
C_{\Lambda^{dt}}(t_a,t_b)  = -{\beta^2 \over 2}f'(C^{dt}(t_a,t_b))
\label{def5}
\eeq
\beq
\hat{R}_{1,\Lambda^{dt}}(t_a,t_b)  = -{\beta^2 \over 2}f''(C^{dt}(t_a,t_b))\, \hat{R}^{dt}_{1}(t_a,t_b)
\label{def6}
\eeq
\beq
\hat{R}_{2,\Lambda^{dt}}(t_a,t_b)  = -{\beta^2 \over 2}f''(C^{dt}(t_a,t_b))\, \hat{R}^{dt}_{2}(t_a,t_b)
\label{def7}
\eeq
\beqd
\hat{X}_{\Lambda^{dt}}(t_a,t_b)  = -{\beta^2 \over 2}f''(C^{dt}(t_a,t_b))\, \hat{X}^{dt}(t_a,t_b)+
\eeqd
\beq
-{\beta^2 \over 2}f'''(C^{dt}(t_a,t_b))\, \hat{R}^{dt}_{1}(t_a,t_b)\hat{R}^{dt}_{2}(t_a,t_b)\ .
\label{def8}
\eeq
The operator $\Gamma(ab)$ is defined through
\beq
{1 \over 2} \int da db \, \Gamma(ab)q(a)q(b) \equiv \int {1 \over \Gamma_0}\left({\dot{q}^2 \over 4}-\hat{x}^2 \right) dt \  ,
\eeq
that leads to
\beq
\Gamma(ab) \equiv {1 \over \Gamma_0}\left[-{1 \over 2}\delta''(t_a-t_b)\eta_a \, \eta_b-2\, \delta(t_a-t_b) \right]\ .
\label{Gamma12}
\eeq
For $\Gamma_0=1$ we have:
\beqa
\Gamma(ac)Q(cb) & = & -2 \hat{R}_1(t_a,t_b)-{1 \over 2}{d^2 \over dt_a^2}C(t_a,t_b)\eta_a+
\nonumber
\\
& - & 2 \hat{X}(t_a,t_b)\eta_b-{1 \over 2}{d^2 \over dt_a^2}\hat{R}_2(t_a,t_b)\eta_a \eta_b\  \ .
\nonumber
\eeqa
The corresponding expression in eq. (\ref{eqdt}) is obtained replacing $Q(ab)$ with $Q^{dt}(ab)$.
The term depending on $\mu$ can be written as:
\beqa
\mu(a)Q(ab) & = & \mu(t_a)C(t_a,t_b)+
\nonumber
\\
& + & \eta_a \left( \mu(t_a)\hat{R}_1(t_a,t_b)+\hat{\mu}(t_a)C(t_a,t_b) \right)+
\nonumber
\\
& + &\nonumber \eta_b \,  \mu(t_a) \hat{R}_2(t_a,t_b)+
\\
& + & \eta_a \eta_b \left(  \hat{\mu}(t_a)\hat{R}_2(t_a,t_b)+\mu(t_a)\hat{X}(t_a,t_b) \right) .
\nonumber
\eeqa
As above, the corresponding expression in eq. (\ref{eqdt}) is obtained replacing $Q(ab)$ with $Q^{dt}(ab)$.
In order to complete the derivation of the saddle-point equations we need the expression for a product of the form $A(ac)B(cb)$:
\begin{widetext}
\beqa
A(ac)B(cb) &= & C_A(t_a,t_c)\hR_{1,B}(t_c,t_b)+\hR_{2,A}(t_a,t_c)C_B(t_c,t_b)+
 \eta_a \left(\hat{R}_{1,A}(t_a,t_c)\hR_{1,B}(t_c,t_b)+\hX_A(t_a,t_c)C_B(t_c,t_b) \right)+
\nonumber
\\
& + & \eta_b \left(\hat{R}_{2,A}(t_a,t_c)\hR_{2,B}(t_c,t_b)+C_A(t_a,t_c)\hX_B(t_c,t_b) \right)+
\nonumber
\\
& + & 
 \eta_a \eta_b \left(\hat{R}_{1,A}(t_a,t_c)\hX_{B}(t_c,t_b)+\hX_A(t_a,t_c)\hR_{2,B}(t_c,t_b) \right)\ .
\label{prodab}
\eeqa
For the contributions of the initial and final configurations  in the interaction term we have:
\beqa
\Lambda(a1)Q(1b)+\Lambda(a2)Q(2b)& = &  C_\Lambda(t_a,-\Tau)C(-\Tau,t_b)+C_\Lambda(t_a,\Tau)C(\Tau,t_b)
\nonumber
\\
& + &  \eta_a \left(  
\hR_{1,\Lambda}(t_a,-\Tau)C(-\Tau,t_b)+\hR_{1,\Lambda}(t_a,\Tau)C(\Tau,t_b)
 \right)+
\nonumber
\\
& + & \eta_b \left(     C_{\Lambda}(t_a,-\Tau)\hR_2(-\Tau,t_b)+ C_{\Lambda}(t_a,\Tau)\hR_2(\Tau,t_b) \right)+
\nonumber
\\
& + & \eta_a \eta_b \left(   \hR_{1,\Lambda}(t_a,-\Tau) \hR_2(-\Tau,t_b) + \hR_{1,\Lambda}(t_a,\Tau) \hR_2(\Tau,t_b) \right)\ .
\eeqa
\end{widetext}
We also have:
\beq
\delta(ab)=(\eta_a+\eta_b)\delta(t_a-t_b)\ .
\label{identityab}
\eeq
Collecting the various components in eq. (\ref{eq}) and (\ref{eqdt}) we obtain eight integro-differential equations that will be written in expanded form in appendix \ref{sec:expanded} and that were written in compact form in section \ref{sub:compact}.
The extremization of expression (\ref{logtranspher}) with respect to $\mu(t)$ and $\hat{\mu}(t)$ gives the following conditions:
\beq
C(t,t,)=1 \, , \ \ \hR_1(t,t)=\hR_2(t,t')={1 \over 2}\ .
\eeq
note that the last term in  (\ref{logtranspher})  is essential to obtain the correct expression for $\hR_1(t,t)$ and $\hR_2(t,t)$

\begin{widetext}

\section{The equations in expanded form}
\label{sec:expanded}
 Expressions (\ref{spequa}) and (\ref{spequadt}) are compact and useful for a numerical treatment. They correspond to eight integro-differential equations that we write in the following in explicit form for completeness. They have to be supplemented with the definitions (\ref{def1}-\ref{def8}) and the boundary conditions discussed in section \ref{sub:compact}.

\beqa
0 & = & -2 \hR_1(t,t')+\mu(t)C(t,t') +
\nonumber
\\
& + &\int_{-\Tau}^{+\Tau} \left(C_\Lambda(t,t'')\hR_{1}(t'',t')+\hR_{2,\Lambda}(t,t'')C(t'',t')\right)dt'' +
\nonumber
\\
& + &(n-1)\int_{-\Tau}^{+\Tau} \left(C_{\Lambda^{dt}}(t,t'')\hR_{1}^{dt}(t'',t')+\hR_{2,{\Lambda^{dt}}}(t,t'')C^{dt}(t'',t')\right)dt'' +
\nonumber
\\
& + & C_\Lambda(t,-\Tau)C(-\Tau,t')+C_\Lambda(t,\Tau)C(\Tau,t') \ .
\label{exp1}
\eeqa

\beqa
\delta(t-t') & = & -{1 \over 2}{d^2 \over dt^2} C(t,t')+ \mu(t)\hR_1(t,t')+\hat{\mu}(t)C(t,t') +
\nonumber
\\
& + &\int_{-\Tau}^{+\Tau} \left(\hat{R}_{1,\Lambda}(t,t'')\hR_{1}(t'',t')+\hX_\Lambda(t,t'')C(t'',t') \right)dt''+
\nonumber
\\
& + &(n-1) \int_{-\Tau}^{+\Tau} \left(\hat{R}_{1,\Lambda^{dt}}(t,t'')\hR_{1}^{dt}(t'',t')+\hX_{\Lambda^{dt}}(t,t'')C^{dt}(t'',t') \right)dt''+
\nonumber
\\
& + &  \hR_{1,\Lambda}(t,-\Tau)C(-\Tau,t')+\hR_{1,\Lambda}(t,\Tau)C(\Tau,t')\ .
\label{exp2}
\eeqa

\beqa
\delta(t-t') & = & -2 \hX(t,t')+\mu(t)\hR_2(t,t') +
\nonumber
\\
& + &\int_{-\Tau}^{+\Tau}  \left(\hat{R}_{2,\Lambda}(t,t'')\hR_{2}(t'',t')+C_\Lambda(t,t'')\hX(t'',t') \right)dt''+
\nonumber
\\
& + &(n-1) \int_{-\Tau}^{+\Tau}  \left(\hat{R}_{2,\Lambda^{dt}}(t,t'')\hR_{2}^{dt}(t'',t')+C_{\Lambda^{dt}}(t,t'')\hX^{dt}(t'',t') \right)dt''+
\nonumber
\\
& + &     C_{\Lambda}(t,-\Tau)\hR_2(-\Tau,t')+ C_{\Lambda}(t,\Tau)\hR_2(\Tau,t')\ .
\label{exp3}
\eeqa

\beqa
0 & = & -{1 \over 2}{d^2 \over dt^2} \hR_2(t,t')+ \mu(t)\hX(t,t')+\hat{\mu}(t)\hR_2(t,t')+
\nonumber
\\
& + &  \int_{-\Tau}^{+\Tau} \left(\hat{R}_{1,\Lambda}(t,t'')\hX(t'',t')+\hX_\Lambda(t,t'')\hR_{2}(t'',t') \right)dt'' +
\nonumber
\\
& + &(n-1)  \int_{-\Tau}^{+\Tau} \left(\hat{R}_{1,\Lambda^{dt}}(t,t'')\hX^{dt}(t'',t')+\hX_{\Lambda^{dt}}(t,t'')\hR_{2}^{dt}(t'',t') \right)dt'' +
\nonumber
\\
& + &  \hR_{1,\Lambda}(t,-\Tau) \hR_2(-\Tau,t') + \hR_{1,\Lambda}(t,\Tau) \hR_2(\Tau,t')  \ .
\label{exp4}
\eeqa

%
%
%

\beqa
0 & = & -2 \hR_1^{dt}(t,t')+\mu(t)C^{dt}(t,t') +
\nonumber
\\
& + &\int_{-\Tau}^{+\Tau} \left(C_{\Lambda^{dt}}(t,t'')\hR_{1}(t'',t')+\hR_{2,{\Lambda^{dt}}}(t,t'')C(t'',t')\right)dt'' +
\nonumber
\\
& + &\int_{-\Tau}^{+\Tau} \left(C_{\Lambda}(t,t'')\hR_{1}^{dt}(t'',t')+\hR_{2,{\Lambda}}(t,t'')C^{dt}(t'',t')\right)dt''  +
\nonumber
\\
& + &(n-2)\int_{-\Tau}^{+\Tau} \left(C_{\Lambda^{dt}}(t,t'')\hR_{1}^{dt}(t'',t')+\hR_{2,{\Lambda^{dt}}}(t,t'')C^{dt}(t'',t')\right)dt'' +
\nonumber
\\
& + & C_\Lambda(t,-\Tau)C(-\Tau,t')+C_\Lambda(t,\Tau)C(\Tau,t') \ .
\label{exp1dt}
\eeqa

\beqa
0  & = & -{1 \over 2}{d^2 \over dt^2} C^{dt}(t,t')+ \mu(t)\hR_1^{dt}(t,t')+\hat{\mu}(t)C^{dt}(t,t') +
\nonumber
\\
& + & \int_{-\Tau}^{+\Tau} \left(\hat{R}_{1,\Lambda^{dt}}(t,t'')\hR_{1}(t'',t')+\hX_{\Lambda^{dt}}(t,t'')C(t'',t') \right)dt''+
\nonumber
\\& + &\int_{-\Tau}^{+\Tau} \left(\hat{R}_{1,\Lambda}(t,t'')\hR_{1}^{dt}(t'',t')+\hX_{\Lambda}(t,t'')C^{dt}(t'',t') \right)dt''+
\nonumber
\\
& + &(n-2) \int_{-\Tau}^{+\Tau} \left(\hat{R}_{1,\Lambda^{dt}}(t,t'')\hR_{1}^{dt}(t'',t')+\hX_{\Lambda^{dt}}(t,t'')C^{dt}(t'',t') \right)dt''+
\nonumber
\\
& + &  \hR_{1,\Lambda}(t,-\Tau)C(-\Tau,t')+\hR_{1,\Lambda}(t,\Tau)C(\Tau,t')\ .
\label{exp2dt}
\eeqa

\beqa
0 & = & -2 \hX^{dt}(t,t')+\mu(t)\hR_2^{dt}(t,t') +
\nonumber
\\
& + & \int_{-\Tau}^{+\Tau}  \left(\hat{R}_{2,\Lambda^{dt}}(t,t'')\hR_{2}(t'',t')+C_{\Lambda^{dt}}(t,t'')\hX(t'',t') \right)dt''+
\nonumber\\
& + & \int_{-\Tau}^{+\Tau}  \left(\hat{R}_{2,\Lambda}(t,t'')\hR_{2}^{dt}(t'',t')+C_{\Lambda}(t,t'')\hX^{dt}(t'',t') \right)dt''+
\nonumber
\\
& + &(n-2) \int_{-\Tau}^{+\Tau}  \left(\hat{R}_{2,\Lambda^{dt}}(t,t'')\hR_{2}^{dt}(t'',t')+C_{\Lambda^{dt}}(t,t'')\hX^{dt}(t'',t') \right)dt''+
\nonumber
\\
& + &     C_{\Lambda}(t,-\Tau)\hR_2(-\Tau,t')+ C_{\Lambda}(t,\Tau)\hR_2(\Tau,t')\ .
\label{exp3dt}
\eeqa

\beqa
0 & = & -{1 \over 2}{d^2 \over dt^2} \hR_2^{dt}(t,t')+ \mu(t)\hX^{dt}(t,t')+\hat{\mu}(t)\hR_2^{dt}(t,t')+
\nonumber
\\
& + &  \int_{-\Tau}^{+\Tau} \left(\hat{R}_{1,\Lambda^{dt}}(t,t'')\hX(t'',t')+\hX_{\Lambda^{dt}}(t,t'')\hR_{2}(t'',t') \right)dt'' +
\nonumber
\\& + &  \int_{-\Tau}^{+\Tau} \left(\hat{R}_{1,\Lambda}(t,t'')\hX^{dt}(t'',t')+\hX_{\Lambda}(t,t'')\hR_{2}^{dt}(t'',t') \right)dt'' +
\nonumber
\\
& + &(n-2)  \int_{-\Tau}^{+\Tau} \left(\hat{R}_{1,\Lambda^{dt}}(t,t'')\hX^{dt}(t'',t')+\hX_{\Lambda^{dt}}(t,t'')\hR_{2}^{dt}(t'',t') \right)dt'' +
\nonumber
\\
& + &  \hR_{1,\Lambda}(t,-\Tau) \hR_2(-\Tau,t') + \hR_{1,\Lambda}(t,\Tau) \hR_2(\Tau,t')  \ .
\label{exp4dt}
\eeqa

\end{widetext}

\section{The solutions on the corners and on the diagonal}

In this appendix we discuss a number of useful properties that follow from the equations.
Equation \ref{exp3}  implies that
\beq
 \hX(t,t')=-{1 \over 2}\delta(t-t')+\delta \hX(t,t')
\eeq
where $\delta \hX(t,t')$ is a bounded function.
In the limit $t,t' \rightarrow \pm\Tau$ the interaction part in the equations greatly simplifies because we have 
\beqd
C(-\Tau,t)=C^{dt}(-\Tau,t)\, , \  \hR_1(t,-\Tau)=\hR_1^{dt}(t,-\Tau)\, , 
\eeqd
\beqd
\  \delta\hX(-\Tau,t)=\hX^{dt}(-\Tau,t)\ .
\eeqd
This allows to characterize the order parameters on the corners of the $t,t'$ domain.
In particular one easily sees that for $n=0$ the contributions of the integrals between $-\Tau$ and $\Tau$ cancel in the limit $t,t' \rightarrow -\Tau$ in all equations. In particular eq. \ref{exp1} leads to:
\beq
-1+\mu(-\Tau)-{\beta^2 \over 2}f'(1)=0 \rightarrow \lim_{t \rightarrow \pm \Tau}\mu(t)=\mu_{eq}\ .
\eeq
Thus $\mu(t)$ goes continuously to the equilibrium value as $t\rightarrow \pm \Tau$.
Equations \ref{exp3} and \ref{exp3dt} lead to:
\beq
\delta\hX(\pm \Tau,\pm \Tau)=\hX^{dt}(\pm \Tau,\pm\Tau)={\mu_{eq} \over 4}={1 \over 4}+{\beta^2 \over 8}f'(1)
\eeq
Similarly for $n=0$ all terms depending on $\Lambda$ cancels in the limit $t \rightarrow \mp\Tau, \, \ t' \rightarrow \pm\Tau$ due to the boundary conditions
\beqa
0 & = & C(\pm \Tau, \mp \Tau)=C^{dt}(\pm \Tau, \mp \Tau)
\eeqa
and one obtains from eqs. \ref{exp1} and \ref{exp3}: 
\beqa
0 & = &\hR_1(\pm \Tau, \mp \Tau)=\hR_1^{dt}(\pm \Tau, \mp \Tau)
\\
0 & = &\hR_2(\pm \Tau, \mp \Tau)=\hR_2^{dt}(\pm \Tau, \mp \Tau)
\\
0 & = &\delta \hX(\pm \Tau, \mp \Tau)=\hX^{dt}(\pm \Tau, \mp \Tau)
\eeqa

\section{Derivatives of the logarithm of the rate} 
\label{sub:der}
  
The expression (\ref{logtranspher}) of the logarithm of the  rate in terms of the order parameter obtained in the path integral formulation is divergent as is shown explicitly for the free case in appendix (\ref{sec:trfree}) and its computation requires to go back to the discretized case.
A convenient alternative is to compute the derivative of the logarithm of the rate with respect to the temperature and integrate the result using the knowledge of the infinite temperature limit  given in section \ref{sec:free}.
The partial derivative with respect to the temperature of expression (\ref{logtranspher}) is simply given by $ {\beta \over 2 }{d \over dn}\int d{\bf a}d{\bf b}f({\bf Q(a,b)})$ and it coincides with the total derivative when computed on the solution of the saddle-point equations. Using the formulas for the integrals computed below (appendix \ref{subsub:int}) we obtain:
\begin{widetext}
\beq
{1 \over N}\, {\partial \overline{[\ln \hT]} \over \partial \beta}= {\beta \over 2} \left(  4\int_{-\Tau}^{+\Tau} \hR_{1,f[C]}(t,-\Tau)dt+  \int_{-\Tau}^{+\Tau} \int_{-\Tau}^{+\Tau}  \hX_{f[C]}(t,t')dt dt'+(n-1)\int_{-\Tau}^{+\Tau}\int_{-\Tau}^{+\Tau}  \hX_{f[C^{dt}]}(t,t')dt dt' \right)\ .
\label{devrate}
\eeq 
\end{widetext}
As discussed in sec. \ref{sub:ergres}
 it is interesting to consider a generalized $[\ln \hT]$ in which the final configuration at time $+\Tau$ is selected with the Gibbs weight corresponding to a different temperature $\beta_2$.
This allows to compute the total probability to jump to states with energy higher than the equilibrium one.
Again the actual computation of the divergent expression can be avoided studying the derivative of  $[\ln \hT]$ with respect to $\beta_2$. The same argument above implies that the total derivative coincides with the partial derivative. With a computation similar to that leading to eq. (\ref{intgen}) one finds: 
\beq
{1 \over N}\, {\partial \overline{[\ln \hT]} \over \partial \beta_2}= \frac{\beta}{2} \int_{-\Tau}^{+\Tau} \hR_{1,f[C]}(t,\Tau)dt
\eeq 
Note that when $\beta_2 \neq \beta$ the order-parameter functions are no longer symmetric with respect to the exchange $t \rightarrow -t$.
The probability of jumping to the configurations with energy $E_2$ (the equilibrium energy at inverse temperature $\beta_2$) with typical rate is then:
\beq
\exp\left[ \beta E(\beta)/2 +\overline{[\ln \hT]}  -\beta E(\beta_2)/2+S(\beta_2)  \right]\eeq 
and the derivative of the logarithm of the above expression with respect to $\beta_2$  is:
\beq
{\beta \over 2} \int_{-\Tau}^{+\Tau} \hR_{1,f[C]}(t,\Tau)dt-{\beta \over 4}f(1)\ .
\label{derb2}
\eeq
I have not solved numerically the case $\beta_2\neq \beta$, but the evaluation of the  the derivative at $\beta_2=\beta$ allows to characterize qualitatively the total logarithm of the rate for $\beta_2 \neq \beta$ in the neighborhood of $\beta$.
This quantity turns out to be negative when evaluated on the numerical solution at $\beta=1.695$ with $\beta_2=\beta$ meaning that the total rate is larger considering configurations with energies larger than equilibrium, which is the key result at bases of the discussion of section (\ref{sub:ergres}).

Note that consistently in the ergodic phase expression (\ref{derb2}) vanishes  in the $\Tau \rightarrow \infty$ limit because the solutions satisfy eqs. (\ref{eqR}).
Instead, as we just saw, for $T<T_d$ the above quantity is negative for $\beta_2=\beta$ also in the $\Tau \rightarrow \infty$ limit and should vanish for some $\beta_2=\beta_{max}<\beta$  corresponding to a maximum at a higher energy (and temperature) with $\beta_{max} \rightarrow \beta$ as $T \rightarrow T_d^-$ .

\subsection{Integrals }
\label{subsub:int}
We consider the expression of  the sum of the elements of a generic object $\bm{B(ab)}$ in terms of its components,
\beq
\int \bm{B(ab) da db} \ .
\eeq
We specialize to the case in which $\bm{B(ab)}$ has the structure of the order parameter $\bm{Q(ab)}$ as discussed in section \ref{sec:action}.
In particular we have for the $m+m'$ static replicas:
\beq
B(\alpha,\beta)=\delta_{\alpha\beta} C_B(\Tau,\Tau)
\eeq 
similarly the $n$ dynamical replicas are correlated only with the static replicas associated to the initial and final conditions and their correlations have a RS form.
We define
\beq
\bm{A(a)} \equiv \int \bm{B(ba)db}\ .
\eeq 
For $\bm a$ corresponding to any of the $m+m'$ static replicas other than those fixing the initial and final condition we simply have 
\beq
\bm{A(a)}=C_B(\Tau,\Tau)
\eeq
For $\bm a$ corresponding to either one of the two static replicas controlling the initial and final conditions we have:
\beqa
\bm{A}(\Tau) & = & \int \bm{B}(\Tau,\bm{b})\bm{db}=C_B(\Tau,\Tau)+C_B(\Tau,-\Tau)+
\nonumber
\\
& + & 
n \int \hR_{2,B}(\Tau,t)dt
\eeqa
For $\bm a$ given by the $\eta$ component of one of the dynamical replicas we have:
\beqa
\hat{a}(t) & = & \hR_{1,B}(t,\Tau)+\hR_{1,B}(t,-\Tau)+\int \hX_B(t,t')dt'+
\nonumber
\\
& + & 
(n-1)\int \hX_B^{dt}(t,t')dt' \, .
\eeqa
Putting everything together one finds:
\begin{widetext}
\beqa
\int \bm{B(ab) da db} &  = & \int \bm{A(a) da }= (m+m')\, C_B(\Tau,\Tau)+
\nonumber
\\
& + &n \left(  \int \hR_{1,B}(t,-\Tau)dt+\int \hR_{1,B}(t,\Tau)dt  +\int \hR_{2,B}(\Tau,t)dt+\int \hR_{2,B}(-\Tau,t)dt\right. +
\nonumber
\\
& + & \left. \int  \hX_B(t,t')\,dt \,dt'+(n-1)\int \hX_{B^{dt}}(t,t')\,dt\, dt' \right)
\label{intgen}
\eeqa
\end{widetext}

\section{The energy} 

The instantaneous energy on the trajectory reads:
\beq
E(t)\equiv \sum_{p=1}^{\infty} \sum_{i_1 < \dots < i_p} \overline{ J_{i_1\dots i_p}[\langle s_{i_1}(t)\dots s_{i_p}(t) \rangle ]}\ .
\eeq
Exploiting the fact that the $J$'s are Gaussian random variables through an integration by part one easily obtains:
\begin{widetext}
\beq
e(t)\equiv {E(t) \over N}=-{\beta  \over 2}\left(  C_{f[C]}(t,-\Tau)+C_{f[C]}(t,\Tau)  +  \int_{-\Tau}^{+\Tau}  \hR_{2,f[C]}(t,t') dt'+(n-1)\int_{-\Tau}^{+\Tau} \hR_{2,f[C^{dt}]}(t,t') dt'\right)
\label{ene}
\eeq
\end{widetext}
In the special case of the pure $p$-spin {\it i.e.} $f(x)=x^p$ this leads to a simple relationship between the energy and $\mu(t)$:
\beq
e(t)={1 \over p\, \beta}(1-\mu(t))\ .
\eeq
The above can be shown using equation (\ref{exp1}) at equal times and using the property that 
\beq
f'(x)+f''(x)\, x=p \,f'(x)\, ,\  \mathrm{for}\ \ f(x)=x^p\ ,
\eeq
to make a connection with expression (\ref{ene}).

\section{The Transition Rate in The free case}
\label{sec:trfree}

In the free case the expression for the Replicated logarithm transition rate (\ref{logtranspher}) simplifies considerably due to $\beta=0$ and $\bm{\Lambda(ab)}=0$. In the RS ansatz one obtains:
\begin{widetext}
\beqd
n \left(  {1 \over 2 } \int dt \,   \hmu (t)+{ \Gamma_0 \over 2} \int   \mu(t)dt + \int\ {d \sigma \over \sqrt{2 \pi}}e^{-{\sigma^2 \over 2}}\int {d \tau \over \sqrt{2 \pi}}  e^{ -{\tau^2 \over 2}} \ln Z(\sigma,\tau)\right )  \ ,
\eeqd
\beq
Z(\sigma,\tau) = \int_{q(-\Tau)=\sigma}^{ q(\Tau)=\tau} [dq] \exp\left[\int dt\left(  -{\dot{q}^2 \over  4 \, \Gamma_0}-{\Gamma_0 \over 4}\mu^2 q^2-{1  \over 2}\hmu q^2 \right)  \right]\ .
\eeq
\end{widetext}
We recognize the path integral representation of the Harmonic oscillator that is usually written as
\beqd
Z(\sigma,\tau) = \int_{q(-\Tau)=\sigma}^{ q(\Tau)=\tau} [dq] \exp\left[-\int dt\left(  {1 \over 2} \, m \, \dot{q}^2 +{1 \over 2}m \omega^2 q^2 \right)  \right]
\eeqd

with the identification $m=1/(2\Gamma_0)$ that leads to the same $\dot{q}^2$ factor and the same $[dq]$.
As discussed in classic textbooks the above path integral is ill defined. This is easily seen switching to a frequency representation where it is ultraviolet  divergent as $\ln Z \propto \int^{\infty} dk \,  \ln (\omega^2 +k^2)$.
The actual quantity $Z(\sigma,\tau)$ is finite because the differential $[dq]$ includes a prefactor diverging as $1/\Delta t$  as we have seen in section (\ref{sec:pathint}). A careful computation leads to the following expression for $\ln Z(\sigma,\tau)$ (see eq. 2.23 in Zinn-Justin's \cite{Zinn-Justin2002} or eq. 13.45 in Parisi's \cite{Parisi1988} ):
\beqd
{1 \over 2}\ln { m \omega \over 2 \pi \sinh \omega \tau}-{m \omega \over 2 \sinh \omega \tau}\left[ \cosh \omega \tau(\sigma^2+\tau^2 )-2 \sigma \tau \right]
\eeqd
\beqd
m={1 \over 2}\, , \ \omega=\sqrt{a}\, , \ \tau \equiv =t_{fin}-t_{in} = 2 \Tau \, ,
\eeqd
where we have fixed $\Gamma_0=1$ and used the results $\mu=1$ and $a \equiv 1+2 \hmu$ obtained in section \ref{sec:free}.
Performing the averages over $\sigma$ and $\tau$ we finally obtain:
\beq
{1 \over N}\overline{[\ln \hT]}= -{1 \over 2}\ln 2 \pi-{1 \over 2} + (1+\hmu) \Tau +{1 \over 4} \ln {- \hmu \over 2}
\eeq

\bibliography{biblio.bib}

\end{document}